\documentclass[manuscript]{aastex}
\usepackage{glossaries}
\usepackage{savesym}
\savesymbol{tablenum}
\usepackage[range-units=single,list-units=single]{siunitx}
\restoresymbol{SIX}{tablenum}

\usepackage[version=4]{mhchem}

\newcommand{\planet}{Planet 1.0}
\newcommand{\modele}{ModelE2}
\newcommand{\rocke}{ROCKE-3D}
\DeclareSIUnit{\astronomicalunit}{\text{AU}}

\begin{document}

\title{Resolving Orbital and Climate Keys of Earth and Extraterrestrial Environments with Dynamics (ROCKE-3D) 1.0: A General Circulation Model for Simulating the Climates of Rocky Planets}

\author{M.~Way}
\affil{NASA Goddard Institute for Space Studies, New York, NY 10025, USA}
\affil{Department of Physics and Astronomy, Uppsala University, Uppsala, 75120, Sweden}

\author{I.~Aleinov}
\affil{NASA Goddard Institute for Space Studies, New York, NY 10025, USA}
\affil{Center for Climate Systems Research, Columbia University, New York, NY 10025, USA}

\author{David~S.~Amundsen}
\affil{NASA Goddard Institute for Space Studies, New York, NY 10025, USA}
\affil{Department of Applied Physics and Applied Mathematics, Columbia University, New York, NY 10025, USA}

\author{M.~Chandler}
\affil{NASA Goddard Institute for Space Studies, New York, NY 10025, USA}
\affil{Center for Climate Systems Research, Columbia University, New York, NY 10025, USA}

\author{T.~Clune}
\affil{Global Modeling and Assimilation Office, NASA Goddard Space Flight Center}

\author{A.~D.~Del~Genio}
\affil{NASA Goddard Institute for Space Studies, New York, NY 10025, USA}

\author{Y.~Fujii}
\affil{NASA Goddard Institute for Space Studies, New York, NY 10025, USA}

\author{M.~Kelley}
\affil{NASA Goddard Institute for Space Studies, New York, NY 10025, USA}

\author{N.~Y.~Kiang}
\affil{NASA Goddard Institute for Space Studies, New York, NY 10025, USA}

\author{L.~Sohl}
\affil{NASA Goddard Institute for Space Studies, New York, NY 10025, USA}
\affil{Center for Climate Systems Research, Columbia University, New York, NY 10025, USA}

\author{K.~Tsigaridis}
\affil{NASA Goddard Institute for Space Studies, New York, NY 10025, USA}
\affil{Center for Climate Systems Research, Columbia University, New York, NY 10025, USA}

\date{\today}

\begin{abstract}
Resolving Orbital and Climate Keys of Earth and Extraterrestrial Environments
with Dynamics (\rocke) is a 3-Dimensional General Circulation Model (GCM)
developed at the NASA Goddard Institute for Space Studies for the modeling of
atmospheres of Solar System and exoplanetary terrestrial planets. Its parent
model, known as \modele{}  (Schmidt et al. 2014), is used to simulate modern
and 21st Century Earth and near-term paleo-Earth climates. \rocke{} is an
ongoing effort to expand the capabilities of \modele{} to handle a broader
range of atmospheric conditions including higher and lower atmospheric
pressures, more diverse chemistries and compositions, larger and smaller planet
radii and gravity, different rotation rates (slowly rotating to more rapidly
rotating than modern Earth, including synchronous rotation), diverse ocean and
land distributions and topographies, and potential basic biosphere functions.
The first aim of \rocke{} is to model planetary atmospheres on terrestrial
worlds within the Solar System such as paleo-Earth, modern and paleo-Mars,
paleo-Venus, and Saturn's moon Titan. By validating the model for a broad range
of temperatures, pressures, and atmospheric constituents we can then expand its
capabilities further to those exoplanetary rocky worlds that have been
discovered in the past and those to be discovered in the future. We discuss the
current and near-future capabilities of \rocke{} as a community model for
studying planetary and exoplanetary atmospheres.  \end{abstract}

\section{Introduction}\label{sec:introduction}

The rapidly expanding list of confirmed exoplanet detections and accumulating
evidence about the histories of planets in our Solar System has created an
increasing demand for tools that can complement available observations to
provide insights about which planets may be habitable or inhabited, now or in
their past. To date, studies of the climates and habitability of planets other
than modern Earth have been carried out primarily with one-dimensional (1-D)
radiative-convective models
\citep[e.g.][]{Kasting1988,Kasting1993,Pavlov2001,Segura2003,
Domagal-Goldman2008, Domagal-Goldman2011, Kitzmann2010,Zsom2012, Kopparapu2013,
Ramirez2014a, Ramirez2014b, Rugheimer2013, Rugheimer2015, Grenfell2014,
Meadows2016}. These models have the virtue of computational efficiency,
permitting exploration of a wide range of parameter space and coupling to
complex atmospheric chemistry models. Their limitations are their inability to
properly account for the effects of clouds, atmospheric and oceanic heat
transports, obliquity effects, day-night contrasts, and regional aspects of
habitability. 

Modeling of terrestrial climate and climate change was initially performed with
1-D models as well \citep[e.g.][]{Manabe1964, Hansen1981}, but soon gave way to
three-dimensional (3-D) general circulation models (GCMs; sometimes referred to
as global climate models), which are lower resolution versions of the models
used for numerical weather prediction. GCMs have evolved from
atmosphere-only to coupled atmosphere-ocean-sea ice models, and more recently
have added atmospheric and ocean chemistry, land and ocean ecosystem dynamics,
and dynamic land ice to create today’s Earth system models \citep{Jakob2014}
that are the basis of projections of 21st Century anthropogenically forced
climate change.

The first application of a GCM to another planet was the Mars model of
\citet{Leovy1969}, and \citet{Joshi1997} performed the first hypothetical
exoplanet GCM simulation. Since these pioneering studies, GCMs have been used
to simulate the dynamics and climates of a broad range of rocky planets past
and present, as well as planets with thick H$_2$-He envelopes \citep[see][for a
review]{Forget2014}. GCMs self-consistently represent all the processes that
1-D models cannot, though they have their own limitations: uncertainties in
parameterizations of small scale processes, computational cost that requires
radiative transfer and chemistry to be represented in less detail than in 1-D
models, and a level of detail that cannot be constrained as well by
observations for other planets as it can be for Earth. Increasingly, GCMs are
playing a key role in a ``system science'' approach that considers planetary
climate and habitability in the larger context of the evolution of the solid
planet, its parent star, and other planets and planetesimals that affect its
evolution.

In this paper we describe a new planetary and exoplanet GCM, the \rocke{}
(Resolving Orbital and Climate Keys of Earth and Extraterrestrial Environments
with Dynamics) model. \rocke{} is developed from its parent Earth climate GCM,
the NASA Goddard Institute for Space Studies (GISS) \modele{}
\citep{Schmidt2014}. \modele{} was the GISS GCM version used for the Coupled
Model Intercomparison Project Phase 5 (CMIP5), the most recent phase of a
protocol by which successive generations of Earth climate model results are
made publicly available for systematic analysis by the international community.
\rocke{} is configured to simulate the present and past atmospheres of rocky
Solar System planets as well as rocky exoplanets. Like several other planetary
GCMs, \rocke{} is adapted from a previously existing Earth GCM \citep[e.g.
PlanetWRF,][]{Richardson2007}. Unlike any other planetary GCM \rocke{} is based
on the most recent published version of its parent Earth model, is developed
and used in part by the same people who develop the Earth model, and will
evolve in parallel with future generations of the Earth model, thus benefiting
from emerging insights from Earth science into physical processes that are also
relevant to other planets.

The baseline \rocke{} version described in this paper is referred to as
\planet{}. In the following sections we discuss the challenges involved in
adapting an Earth GCM to simulate other rocky planets, the choices made to make
\planet{} as generally applicable as possible, and the remaining limitations
that will not be addressed until the next generation of the model has been
developed. \rocke{} \planet{} has already been used to simulate hypothetical
ancient Venus scenarios \citep{Way2016}, while simulations of several deep
Earth paleoclimate eons, modern Mars, and hypothetical exoplanets are in
progress.

In principle it should be possible to modify an Earth GCM to simulate other
planets simply by changing relevant external parameters. In reality, though,
terrestrial GCMs are designed with only Earth in mind, and are programmed by a
large group of people of varying backgrounds and experience whose composition
evolves over several decades. At any moment in its history, therefore, a GCM is
a mix of modern and obsolete programming approaches, visionary and myopic
coding philosophies, and best and worst practices that necessitate new
approaches to make the model sufficiently general for planetary applications.
Many of those approaches will be discussed herein.

In Section~\ref{sec:configurations} below we discuss the present \planet{}
model resolution and possible ocean configurations. In
Section~\ref{sec:calendar} extensions to the model calendar system are
reviewed. These allow for slower or faster rotating worlds (than present day
Earth), synchronously rotating worlds, and even retrograde rotation like that
of present day Venus. Section~\ref{sec:parameterizations} discusses the major
physics parameterizations in the model, while Section~\ref{sec:properties}
covers its geophysical properties. Section~\ref{sec:enhancement} describes
several examples of GCM modifications for \planet{} that have fed back to the
parent Earth GCM. Section~\ref{sec:use} covers appropriate uses for \rocke, and
Section~\ref{sec:discussion} contains our conclusions. Two appendices provide a
description of input and post processing tools available external to the model.

\section{Model Configurations} \label{sec:configurations}

\subsection{Resolution and Throughput}

In describing the physics of \rocke, we refer to physics from the present
operational version of the parent Earth model as that of ``GISS'' or
``\modele'', and new capabilities as that of ``\rocke''. \modele{} is a
Cartesian gridpoint model routinely run at $\ang{2} \times \ang{2.5}$
latitude-longitude atmospheric resolution with 40 vertical layers, and at
$\ang{1} \times \SI{1.25}{\degree}$ latitude-longitude ocean resolution with 32
vertical layers. This resolution has been retained for certain deep Earth
paleoclimate simulations, where the higher resolution permits better comparison
to geological data as well as better portrayal of the atmospheric and oceanic
dynamics. 

GCM atmospheric (as opposed to oceanic) resolution should at a minimum be fine
enough to crudely resolve the dominant scales of atmospheric motion. Typically
this is assessed using the Rossby radius of deformation (the typical spatial
scale of midlatitude low and high pressure centers) $L_d=NH/f$, where $N$, the
Brunt–V\"{a}is\"{a}l\"{a} frequency, is proportional to the static stability,
$H$, the scale height, depends on temperature, gravity, and atmospheric
composition, and $f$, the Coriolis frequency, is proportional to planet
rotation rate. For Earth $L_d \sim \SI{1000}{\kilo \meter}$ ($\sim 1/6$ Earth's radius) and
$\ang{2} \times \ang{2.5}$ grid boxes are about \SIrange{200}{250}{\kilo \meter} in size, allowing
such features to be adequately resolved. For simulations of other planets, most
initial studies with \planet\ have been for smaller planets for which grid
boxes at the same resolution are smaller or more slowly rotating planets for
which $L_d$ is larger than on Earth. For these simulations it has been possible
to run \planet{} at $\ang{4}\times\ang{5}$ atmospheric and oceanic horizontal
resolution with no loss in accuracy but at almost an order of magnitude faster
speed. This lower resolution version of \planet{} has 20 atmospheric layers
(but with an option for 40 layers) with a model top at \SI{0.1}{\hecto \pascal}
(about \SI{60}{\kilo \meter} altitude), and in coupled mode, 13 ocean layers with maximum depth
up to \SI{4647}{\meter}.
 
\planet{} can be run on a capable laptop for modest integrations at this
coarser resolution, but the bulk of our research is conducted on the NASA
Goddard Space Flight Center Discover cluster of Linux scalable units
(\url{https://www.nccs.nasa.gov/services/discover}). With 44 cores, \rocke{}
can simulate 100 years in approximately 24 hours of wall-clock time with a
fully-coupled ocean at an atmosphere and ocean resolution of $\ang{4} \times
\ang{5}$ with 40 atmospheric layers and 13 ocean layers, using the default
\modele{} radiation scheme. These simulations use a single node/motherboard
with two Intel Xeon E5-2697 v3 Haswell \SI{2.6}{\giga \hertz} each with 14
cores. With SOCRATES, our new radiation scheme (see Section
\ref{sec:radiation}), with the default present day Earth setup we can simulate
approximately 100 years in 48 hours of wall-clock time using 44 cores on the
same cluster.

The parameterized physics in \planet{} is largely the same as that in \modele{},
but several changes that were made after \citet{Schmidt2014} to correct ocean
and radiation physics errors have been adopted for \planet{}.

\subsection{Ocean Models}\label{subsec:oceanmodel}

The oceans are crucial to the accurate 4-D portrayal of a planet's climate
system. Energy, moisture and momentum are exchanged between the atmosphere and
oceans, and the transitions between different phases of water drive some of the
most significant feedback mechanisms operating in the climate system. The
oceans provide the major source of moisture that drives the hydrological cycle,
while the freezing and melting of surface waters have a major impact on
planetary albedo. Together with the transport of heat, these atmosphere-ocean
interactions affect the geographic, seasonal, inter-annual and even
geologic-scale variations of a planet's climate. In \planet{} the oceans differ
from other bodies of water (lakes, rivers) in that salinity and temperature
combine to alter the 3-D density structure, while surface wind stress is
allowed to impact movement of water in the upper ocean. Salinity, temperature,
and wind stress drive global ocean currents that transport energy on time
scales that may exceed the orbital period of the planet by orders of magnitude.
Ocean albedo is a function of both water and sea foam reflectance.  Water
albedo is calculated as a function of the solar zenith angle and wind speed;
the sea foam reflectance is derived from \cite{Frouin1996}

\planet{} allows for 3 different modes of ocean interaction. From simplest to
most complex these are 1) specified sea surface temperature (SST), 2)
thermodynamic upper ocean mixed-layer, and 3) coupled dynamic ocean GCM.

\subsubsection{Specified Ocean Surface Conditions}\label{subsubsec:specified-ocean}

Specifying sea surface temperature (SST), including sea ice cover, is a common
Earth climate modeling technique where SST observations are used as a surface
boundary condition over a range of years or months to force an atmospheric GCM
(AGCM). The GISS model uses twelve monthly arrays that define the ocean surface
temperature and sea ice distributions. The model interpolates the input into
daily values, providing smoother transitions through an annual cycle.
Specifying SSTs is the most commonly accepted technique for evaluating the
efficacy of AGCM physics parameterizations when surface conditions are
well-known (e.g., in performing hindcasts of 20th Century climate). It is also
used in Earth paleoclimate studies where proxy data can be used to reconstruct
past ocean temperature distributions \citep[e.g.][]{MARGO2009}. In this case
the purpose is generally to evaluate the consistency of land-based and
ocean-based observations or simply to examine potential states of the
atmosphere for various time periods in Earth history. Specified SST simulations
are also used to collect the atmosphere-ocean flux information to generate the
Q-fluxes to run the model in mixed-layer ocean mode.

\subsubsection{Mixed-Layer (Q-flux) Ocean Model}\label{subsubsec:qflux-ocean}

For other planets, prescribed SSTs are not an option and SSTs must instead be
calculated interactively to be consistent with a given planet's atmosphere and
external forcing.  The simplest way to do this is to couple the AGCM to a
simple thermodynamically active layer that represents the upper well-mixed
layer of the ocean (typically tens to hundreds of meters deep). The temperature
of the mixed layer responds to radiative and turbulent (sensible and latent)
fluxes of heat across the ocean-atmosphere and ocean-sea ice interfaces. This
approach has been the default choice for most exoplanet GCM studies to date
\citep[e.g.][]{Yang2014,Shields2014,Kopparapu2016,Turbet2016}. In the
literature this approach is typically referred to as a “thermodynamic,” “mixed
layer,” “slab,” or “immobile” ocean model. The greatest limitations of this
method are that it neglects horizontal heat transport by ocean currents and
cannot account for deep water formation related to vertical density gradients.

For Earth, where SST observations exist, a variant of the mixed layer approach
known as the ``q-flux'' method has been commonly applied to simulations of
future climates \citep{Miller1983,Russell1985}. In the Q-flux approach, a
control AGCM run with prescribed SSTs is first conducted to define the
radiative and turbulent heat exchanges at the atmosphere-ocean interface that
are consistent with the AGCM’s physics parameterizations. The implied
horizontal ocean heat transport convergences that would be required to produce
the observed SSTs and sea ice cover in each mixed layer gridbox are then
calculated and applied in a second simulation that couples a mixed layer ocean
model to an AGCM as a proxy for the effect of actual ocean heat transports.
Sometimes diffusive heat loss through the lower boundary of the mixed layer is
also included to mimic exchanges of heat with deeper ocean layers that are
otherwise unrepresented in such models.  The implied ocean heat transport
convergences are themselves fixed, but their presence allows for a more
realistic projection of sea ice changes, and thus ice-albedo feedback, in a
changing climate than is possible in a model that completely ignores ocean heat
transport.  Such models have traditionally been used to define the equilibrium
sensitivity of Earth's climate to a doubling of CO$_{2}$ concentration, a
common benchmark for assessing climate model uncertainty.  The q-flux approach
is also unavailable for exoplanet GCM studies, hence their use of purely
thermodynamic ($\text{Q-flux} = 0$) oceans, and \rocke{} includes a $\text{Q-flux}=0$ ocean
option, but the error induced by ignoring ocean heat transport must be kept in
mind in assessing such studies.  An alternative that has been used for
sensitivity studies is to prescribe a latitudinal profile of ocean heat
transport in a mixed layer model with the latitude and magnitude of the peak
transport as free parameters that can be varied \cite[e.g.][]{Rose2015}.
Furthermore, if an existing simulation with a dynamic ocean (see Section
\ref{subsubsec:dynamic-ocean}) is available, the ocean heat transports from
this model can in principle be used as a specified input to an otherwise
thermodynamic ocean model \citep[e.g.][]{Fiorella09012017}. 

\subsubsection{Dynamic Coupled Ocean}\label{subsubsec:dynamic-ocean}

Given the limitations of Q-flux models, recent generations of Earth climate
models have instead coupled more computationally expensive but more realistic
dynamic ocean GCMs (OGCM) to AGCMs to simulate climate change.  Most exoplanet
GCM studies have eschewed the use of OGCMs because of the large thermal inertia
of the ocean and thus the long integration times required to reach equilibrium,
but several studies have revealed the importance of interactive ocean heat
transport to climates of planets in parameter settings very different from that
of Earth \citep{Vallis-Farneti2009,Cullum2014}.  The most dramatic example of
ocean heat transport effects in the exoplanet context is the difference between
the concentric ``eyeball Earth'' open ocean region simulated beneath the
substellar point of a synchronously rotating aquaplanet with a thermodynamic
ocean \citep{Pierrehumbert2011} and the asymmetric ``lobster'' ocean pattern
produced when a dynamic ocean is used \citep{Hu2014}.

The exploration of parameter space for salty-water-ocean composition differs
from that for atmospheric composition in that the former has a more direct
effect on density structure, circulation, and heat transport.  The Earth's
thermohaline circulation was recently placed into perspective by
\cite{Cullum2016} who demonstrated that an increase in mean salinity can cause
the haline component to dominate.

Most ROCKE-3D simulations couple a dynamic ocean to the atmospheric model. The
standard configuration uses a 4$\arcdeg\times5\arcdeg$ resolution with 13 ocean
layers, which decreases model throughput by $\sim \SI{10}{\percent}$ or less compared to a
thermodynamic ocean but increases the equilibration time of the climate from
decades to centuries of simulated time, with the exact time depending on the
assumed ocean depth.  Some of our deep Earth paleoclimate studies instead use
the same $\ang{1}\times \ang{1.25}$ resolution, 32 layer ocean that is used
by \modele.  Transport by unresolved mesoscale eddies is represented by a
unified Redi/GM scheme \citep{Redi1982, Gent1990, Gent1995, Visbeck1997},
as in \modele{}.  The version used by \cite{Schmidt2014} contained a
miscalculation in the isopycnal slopes that led to spurious heat fluxes across
the neutral surfaces, resulting in an ocean interior that was generally too
warm and southern high latitudes that were too cold.  A correction to resolve
this problem was implemented for \modele{}, Earth paleoclimate studies
\citep{Chandler2013}, and is also used by \rocke{}.  The new code uses a
mesoscale diffusivity of \SI{600}{\meter \squared \per \second},
although some ROCKE-3D exoplanet simulations have used a value of
1200 \SI{1200}{\meter \squared \per \second} instead.  The
applicability of mesoscale eddy parameterizations designed for Earth models has
not yet been investigated for planets with different rotation rates and thus
different dominant spatial scales of eddies \citep{Cullum2014}.

\section{Calendar changes for modeling other planets} \label{sec:calendar}

\modele{} uses a clock and calendar to coordinate model operations that are not
active during every time step and to manage binning/averaging for seasonal and
higher-frequency diagnostics. Prior to the development of \planet{}, this
system made assumptions that were incorrect or inconvenient outside the context
of modern Earth. For instance, the number of days per month was hardwired for a
quasi-Julian 365 day calendar. The system did permit varying the rotational and
orbital periods as well as other orbital parameters (obliquity, eccentricity,
and solar longitude), but provided only limited means to relate these to
seasons. Further, a number of model components possessed implicit (hardwired)
constants appropriate to the lengths of modern Earth day and year.

To enable the study of exoplanets the calendar and indeed the entire
time-management system in \modele{} have been been redesigned to be extensible
and highly encapsulated. The latter was crucial to reduce the likelihood of
accidentally reintroducing assumptions about modern Earth into the model by
subsequent developers. The design of this new time management system reflects
the needs and priorities of climate scientists in several respects. The first
priority was to ensure that the default behavior replicates the original
behavior for modern day Earth-based simulations. The other priority was for the
new calendar to preserve, as much as possible, correspondence between planetary
seasons, months, and days with that of Earth in terms of basic orbital
characteristics. Note that other communities have designed planetary calendars
(primarily for Mars) with quite different priorities such as preserving the
number of days per month and the number of seconds per hour \citep{Allison1997,
Allison2000, Gangale1986, Gangale1997, Gangale2005}. Here the priority is to
simplify interpretation of seasonal and diurnal diagnostics, and is similar to
the approach in \cite{Richardson2007}. In particular, the new calendar system
preserves the intuitive notion of the diurnal cycle being divided into 24 equal
``hours'' as well as the seasonal cycle being divided into 12 ``months.'' Note
that a model ``hour'' will therefore not generally be 3600 seconds in duration,
and months can be significantly longer or shorter than 30 days. Additional
machinery minimally tweaks the orbital period and model timestep to ensure that
the simulation has an integral number of time steps per ``day'' and an integral
number of days per year. All times and time intervals are expressed using exact
integer arithmetic to eliminate issues related to numerical roundoff. We thus
guarantee an exact number of simulation time-steps per day and an exact number
of days per year. 

The specific duration of each calendar month is derived as follows. First, the
solar longitude $\phi_i$, where $i=1,2,\dotsc,12$ is computed for the beginning
of each month in a reference Earth orbit and calendar. The beginning of each
month in the planetary calendar is then determined to have the same solar
longitude angle as for the reference month, subject to rounding to ensure an
integral number of days in each month. To relate the longitudes to
times/durations, the corresponding mean anomaly $M_i$ is computed for the
planetary orbit for the start of each month. $M_i$ can be derived from the
solar longitude by standard Keplerian orbit formulae. The starting day-of-year
$d_i$ for each month is then computed by scaling the delta mean anomaly ($M_i -
M_1)$ by the number of calendar days per radian and rounding to the nearest
day:
\begin{equation}
d_i = 1 + \lfloor (M_i - M_1) \frac{ N_d^{\mbox{cal}}}{2 \pi} + \frac{1}{2}\rfloor
\end{equation}
By default, the system uses the model's standard Earth-based orbit and
pseudo-Julian calendar as the reference. Thus, the planetary ``February'' will
tend to be shorter than average simply due to the short duration of February in
the conventional Earth calendar.

Our basic design is to have the system derive an appropriate calendar directly
from the orbital parameters of a given planet. By introducing software
abstractions for both the orbit and the calendar, the system provides a natural
mechanism for further extension. For example, researchers could easily
introduce a leap day system for their favorite exoplanet by creating a new
Fortran module and adding a control hook in the model initialization. This
approach was quite useful as requests for extensions to the basic planetary
calendar arose almost immediately after deployment.

There is a crucial aspect of Earth's orbit that is not particularly generic - a
large separation of scale between days and years such that the number of days
per year is much, much larger than 1. In terms of the conventional
Julian/Gregorian calendars, this permits months to have an integral number of
days while simultaneously having roughly uniform duration. It also allows
climate models to safely ignore fractional remainders of days that lead to
leap-years. However, for extreme orbits, a lack of this separation of scale can
can have spectacular consequences. The number of days per year can be less than
the number of months, and each day can be longer than a year (e.g., modern
Venus). In such cases, the default for our calendar is to break the
correspondence between the calendar day and the solar day and constrain
calendar to have at least 120 calendar days, i.e., at least 10 calendar days
per month on average. The system has runtime switches that can eliminate this
constraint, as well as the constraint that the rotational period is
commensurate with the orbital period. The latter is crucial to differentiate an
orbit such as that of modern Venus from a tidally locked orbit - both of which
are of interest to \rocke{} modelers. Of course, one must exercise extreme
caution when interpreting model diagnostics in such cases. Some months (and
even some years!) may have 0 solar days. A quantity averaged over one season or
even one year may be highly biased as parts of the planet remain entirely day
or entirely night.

For tidally-locked planets, it is convenient to have a mechanism to vary the
longitude of the subsolar point.  For example, \citet{Turbet2016} point out
that for a synchronously rotating world the continents may be concentrated at
either the substellar or anti-stellar point.   This variation is supported in
our framework by the ``hourAngleOffset'' parameter, which controls placement
the continents for a synchronously rotating world at any angle with respect to
the substellar point.  This approach is much simpler than the equivalent shift
of all boundary condition data (topography, etc.).

\begin{figure}
    \centering
    \includegraphics[width=0.45\textwidth]{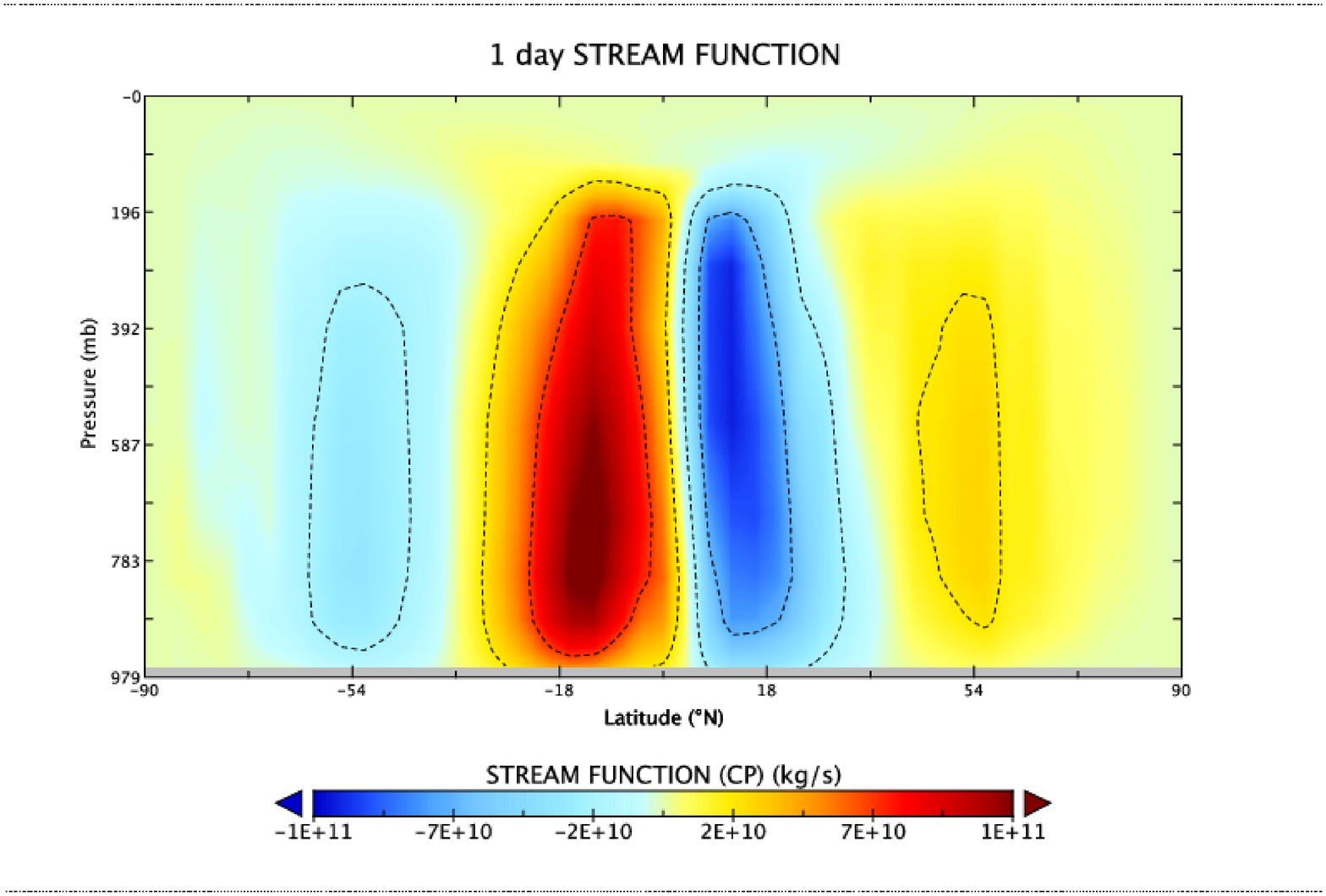}
    \includegraphics[width=0.45\textwidth]{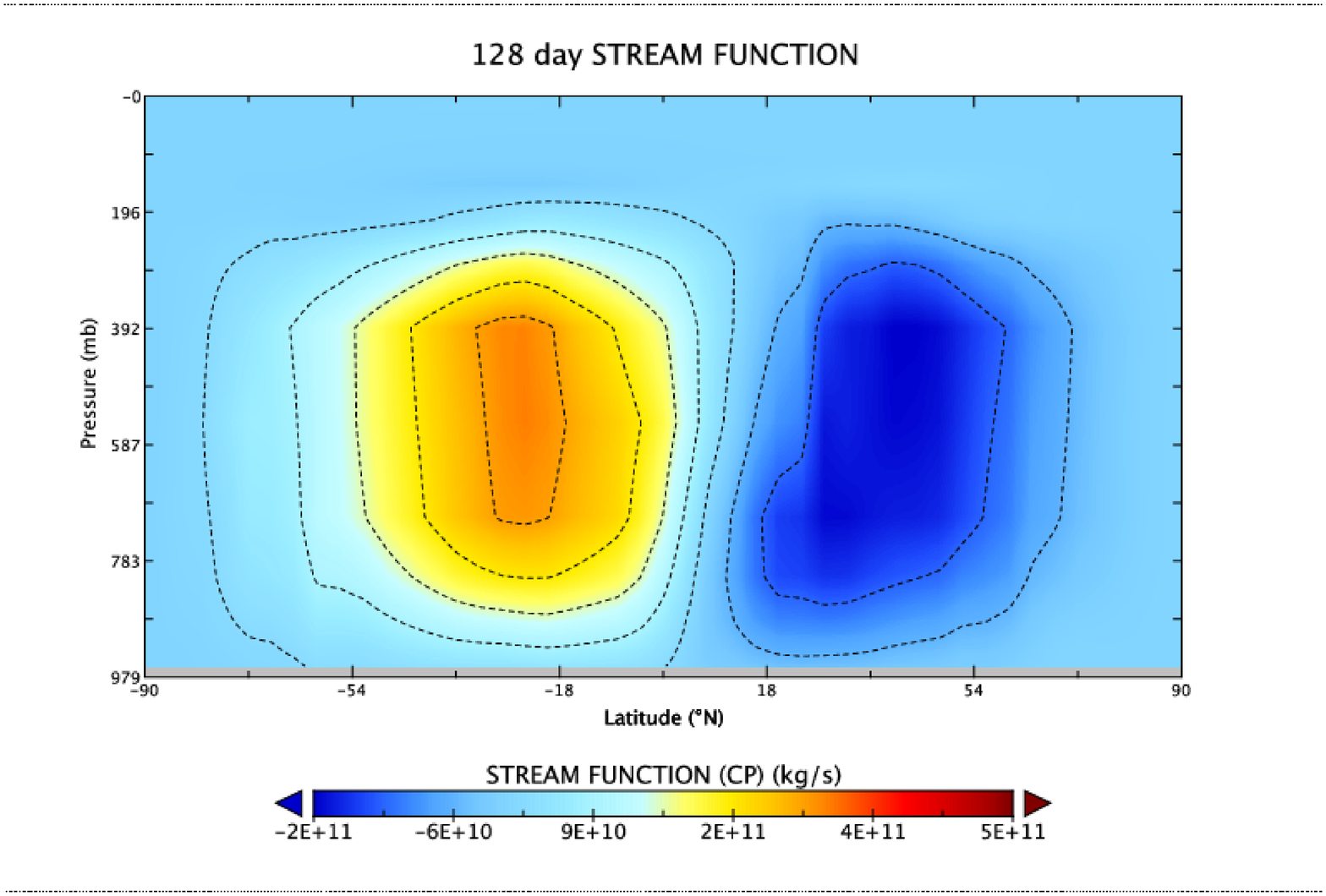}
    \caption{Left: Pressure vs. latitude streamfunction of the mean meridional
circulation of a planet with Earth's rotation period. Right: As in the left
panel but for a planet with a rotation period 128 days longer than an Earth
sidereal day. As expected for a slowly rotating world the Hadley cells are now
much larger in latitudinal extent due to the decrease in the strength of the
Coriolis force at these slow rotation rates. 
}\label{fig:hadley}
\end{figure}

Figure~\ref{fig:hadley} demonstrates that the model responds correctly to the
calendar modifications for slowly rotating worlds as the Hadley cells are
clearly broadened for the slowly rotating planet versus the rapidly (Earth day
length) rotating one.

The calendar has also been expanded to handle variable orbital eccentricities
in time \citep{Way2017}. This would be useful in cases where a Jupiter like
planet perturbs the orbital elements of a nearby smaller terrestrial planet
\citep[e.g.][]{Georgakarakos2016}.

\section{Physics Parameterizations} \label{sec:parameterizations}

Physical processes that operate on scales smaller than those resolved by a GCM
must be parameterized in terms of grid-resolved variables. This section
discusses those which are necessary for \rocke{} and how they are accomplished.

\subsection{Radiation}
\label{sec:radiation}

\subsubsection{The GISS radiation scheme} \label{subsec:radiation_giss}

The radiation scheme in \modele{} was first implemented in \citet{Hansen1983},
with more detailed descriptions of the long-wave radiation scheme in
\citet{Lacis1991} and \citet{Oinas2001}, and the short-wave scheme in
\citet{Lacis1974}. Minor updates have been made to improve its accuracy
\citep{Schmidt2006, Schmidt2014, Pincus2015}, but its overall structure and
parameterizations remain unchanged.

The long-wave radiation scheme uses a 33~$k$-interval $k$-distribution
parameterization derived through Malkmus band models
\citep{Lacis1991,Oinas2001}, with band model parameters derived by fitting
Malkmus band model transmissions to line-by-line transmissions over a range of
absorber amounts. Resulting opacities are tabulated for 19 pressures between
\SIlist{e-6;1}{\bar}, and 8 temperatures between \SIlist{181;342}{\kelvin}. The
radiative transfer equation is solved using six streams without scattering (by
setting the long-wave asymmetry parameter to unity). Long-wave scattering
effects are included via a parameterized correction to the top cloud emission
and outgoing flux, and a slight enhancement of downwelling radiation from cloud
bottom.

Recently updates have been made to the long-wave radiation scheme to improve
its accuracy for atmospheres that deviate slightly from that of the present day
Earth. The tabulated Planck function has been extended to \SI{800}{\kelvin},
and the gas optical depth table has been updated to enable the major greenhouse
gases in the Earth’s atmosphere (H$_2$O, CO$_2$ and O$_3$) to be replaced with
other gases. The latter enables more accurate calculation of fluxes and heating
for cases such as the Archean Earth, which had no O$_3$ nor O$_2$, but may have
had significantly larger amounts of both CO$_2$ and CH$_4$ than present day
Earth. These updates were recently used in the study of the early climate of
Venus \citep{Way2016}.

The short-wave radiation scheme uses the doubling and adding method to include
the effects of multiple scattering \citep{Peebles1951,Hulst1963} with two
quadrature points \citep{Lacis1974}. Gaseous absorption is parameterized
through analytical expressions for the frequency-integrated absorption as a
function of absorber amounts for each gas. The spectrum is divided into 16
gaseous absorption bands, each with one absorbing gas with a corresponding
analytical function for the optical depth as a function of pressure,
temperature and absorber amount.

Stellar radiation input to the GCM drives both the planetary energy balance and
photochemistry. Stellar spectral irradiance (\SIrange{0.115}{100}{\micro
\meter}) to the top-of-the-atmosphere is provided to the model via an input
file that can be changed for different stars, to drive the energy balance and
to provide UV fluxes for ozone calculations and photolysis rates. A software
utility provided by GISS can be used to format high-resolution stellar spectra
for input to the GCM (see Appendix~\ref{subsec:GCMInputs}). The various modules
of the GCM that utilize this spectral irradiance perform different spectral
partitioning to suit their functions, such as for surface albedo, and for
photosynthesis by plants and phytoplankton. Some solar spectral radiation
assumptions are still hard-coded into the model, such as the broadband
absorbance of water, so users should consult GISS personnel when interpreting
results with alternative stellar spectra. More details are provided below in
Section \ref{subsec:Cryosphere} on the Cryosphere.

We note that for this radiation scheme to be reasonably accurate, gas
concentrations of radiatively active gases, and H$_2$O, CO$_2$ and O$_3$ in
particular, should be within a factor of 10 of present day Earth values
throughout the atmosphere. In addition, the short-wave radiation scheme should
only be used with stellar spectra that are of the same spectral type as the
Sun.

\subsubsection{SOCRATES}\label{subsec:radiation_socrates}

In \rocke{} we require a radiation scheme that can be applied to a wide variety
of planetary atmospheres. Consequently, the ability to easily change spectral
bands, extend the pressure and temperature range of opacity tables, and add and
remove absorbers are imperative. Unfortunately, the parametrizations used by
the radiation scheme in \modele{} prohibit such a generalization. To ease
adaptation of \rocke{} to different atmospheres we have coupled it to the Suite
of Community Radiative Transfer codes based on Edwards and
Slingo\footnote{\url{http://code.metoffice.gov.uk/trac/socrates}}
\citep[SOCRATES,][]{Edwards1996a,Edwards1996b}. This radiation scheme is in
operational use in the UK Met Office Unified Model, has previously been adapted
to hot Jupiters \citep{Amundsen2014,Amundsen2016a}, and is available under a
BSD 3-clause
licence\footnote{\url{https://opensource.org/licenses/BSD-3-Clause}}.
Importantly, SOCRATES allows for changing radiation bands, altering pressures
and temperatures in the opacity tables, and the inclusion of various
combinations of gaseous absorbers with relative ease.

SOCRATES solves the two-stream approximated radiative transfer equation with
multiple scattering for both the short-wave and long-wave components. Several
different two-stream approximations are available, but by default we use the
practical improved flux method version from \citet{Zdunkowski1985} with a
diffusivity $D = 1.66$ for the long-wave component and the original version of
\citet{Zdunkowski1980}, which uses a diffusivity $D = 2$, for the short-wave
component. We note that, unlike the two-stream equations presented in
\citet{Toon1989}, the two-stream equations of \citet{Zdunkowski1985} can be
applied with a variable diffusivity, enabling improved accuracy compared to the
\citet{Toon1989} formulation. In order to improve the representation of the
scattering phase functions with strong forward-scattering peaks delta-rescaling
\citep{Thomas2002} is applied for both components.

Gaseous absorption is parameterized using the correlated-$k$ method
\citep{Lacis1991,Goody1989}, with $k$-coefficients derived using exponential
sum fitting of transmissions \citep{Wiscombe1977}, tabulated as a function of
pressure and temperature. We use the HITRAN 2012 line list \citep{Rothman2013}
to calculate cross sections line-by-line using Voigt profiles and the CAVIAR
water vapour continuum~\citep{Ptashnik2011}. For planets with Earth-like
atmospheres orbiting Sun-like stars we use 9 long-wave and 6 short-wave bands,
those used by the UK Met Office for global atmosphere configuration 7.0
\citep[GA7.0,][]{Walters2017}, given in Tables \ref{tab:long-wave} and
\ref{tab:short-wave}, and tabulate $k$-coefficients on 51 pressures equally
spaced in $\log P$ between \SIlist{e-5;1}{\bar}, and 13 temperatures spaced
linearly in temperature between \SIlist{100;400}{\kelvin}. The bands in Tables
\ref{tab:long-wave} and \ref{tab:short-wave} will need to be changed in order
to improve accuracy for atmospheres with significantly different compositions
or stellar irradiation spectra, see e.g. \citet{Fujii2017} where 29 short-wave
bands were used in order to accurately handle absorption of stellar radiation
by water vapor at near-IR wavelengths for large specific humidities. Additional
physics will need to be added to treat atmospheres with a large amount of
CO$_2$ as the effects of CO$_2$ Rayleigh scattering, self-broadening,
non-Voigtian line wings and continuum absorption are currently not supported.

Overlapping gaseous absorption is treated using equivalent extinction
\citep{Edwards1996b}, although random overlap \citep{Lacis1991} is also
supported. Both of these methods combine $k$-coefficients calculated for each
gas separately on-the-fly. Equivalent extinction relies on having a major
absorber in each band, we use the adaptive equivalent extinction approach
described in \citet{Amundsen2016b} to determine the major absorber in each band
independently for each column, which may also change in time. Pre-mixing of
opacities \citep{Goody1989}, where $k$-coefficients are derived directly for
the gas mixture for a given composition, would result in a faster radiation
scheme, however, it requires new $k$-tables to be derived when gas amounts are
changed \citep{Amundsen2016b}

\begin{table}
\centering
\caption{The long-wave bands adopted for planets with Earth-like atmospheres.
These are the bands used by the UK Met Office for global atmosphere
configuration 7.0 \citep[GA7.0,][]{Walters2017}. Note that bands 3 and 5
contain excluded regions.} \label{tab:long-wave}
\begin{tabular}{|l|l|l|}
\hline
Long-wave band & Wavenumber [\si{\centi \meter^{-1}}] & Wavelength [\si{\micro \meter}] \\
\hline \hline
1              & \numrange{1}{400}                            & \numrange{25}{10000} \\
\hline
2              & \numrange{400}{550}                          & \numrange{18.18}{25} \\
\hline
3              & \numrange{550}{590} and \numrange{750}{800}  & \numrange{12.5}{13.33} and \numrange{16.95}{18.18} \\
\hline
4              & \numrange{590}{750}                          & \numrange{13.33}{16.95} \\
\hline
5              & \numrange{800}{990} and \numrange{1120}{1200} & \numrange{8.33}{8.93} and \numrange{10.10}{12.5} \\
\hline
6              & \numrange{990}{1120}                         & \numrange{8.93}{10.10} \\
\hline
7              & \numrange{1200}{1330}                        & \numrange{7.52}{8.33} \\
\hline
8              & \numrange{1330}{1500}                        & \numrange{6.67}{7.52} \\
\hline
9              & \numrange{1500}{2995}                        & \numrange{3.34}{6.67} \\
\hline
\end{tabular}
\end{table}

\begin{table}
\centering
\caption{The short-wave bands adopted for planets with Earth-like atmospheres
orbiting Sun-like stars. These are the bands used by the UK Met Office for
global atmosphere configuration 7.0 \citep[GA7.0,][]{Walters2017}.}
\label{tab:short-wave}
\begin{tabular}{|l|l|l|}
\hline
Short-wave band & Wavenumber [\si{\centi \meter^{-1}}] & Wavelength [\si{\nano \meter}] \\
\hline \hline
1               & \numrange{31250}{50000} & \numrange{200}{320} \\
\hline
2               & \numrange{19802}{31250} & \numrange{320}{505} \\
\hline
3               & \numrange{14493}{19802} & \numrange{505}{690} \\
\hline
4               & \numrange{8403}{14493}  & \numrange{690}{1190} \\
\hline
5               & \numrange{4202}{8403}   & \numrange{1190}{2380} \\
\hline
6               & \numrange{1000}{4202}   & \numrange{2480}{10000} \\
\hline
\end{tabular}
\end{table}

Rayleigh scattering by air is included, and we have also implemented a new
Rayleigh scattering formulation that calculates the Rayleigh scattering
coefficient consistently with the atmospheric composition in each layer. Water
cloud optical properties are derived using Mie scattering, while the
parametrization of ice crystals is described in \citet{Edwards2007} and is
based on the representation of ice aggregates introduced by \citet{Baran2001}.
Vertical cloud overlap is treated using the mixed maximum-random overlap
assumption (clouds in adjacent layers overlap maximally, while clouds separated
by one or more clear layers overlap randomly).

In order to improve the accuracy of the calculated long-wave and short-wave
fluxes, wavelengths are weighted internally in each band by the Planck function
at \SI{250}{\kelvin} for the long-wave component and by the stellar spectrum
for the short-wave component when deriving $k$-coefficients, aerosol and cloud
optical properties. This is important as our bands are quite broad,
particularly for the short-wave component, and the source function varies
significantly within the bands. Consequently, changing the stellar spectrum
involves computing new $k$-coefficients and cloud and aerosol optical
properties for use in the calculation of short-wave fluxes.

As stated in Section~\ref{sec:configurations}, with the GISS radiation scheme
we are able to simulate 100 Earth years in approximately 24 hours of wall-clock
time using 44 cores\footnote{These simulations use a single node/motherboard
with two Intel Xeon E5-2697 v3 Haswell \SI{2.6}{\giga \hertz} each with 14
cores.} with a fully-coupled ocean at an atmosphere and ocean resolution of
$\ang{4} \times \ang{5}$ with 40 atmospheric layers and 13 ocean layers. With
SOCRATES, \rocke{} is significantly slower, and allows us to simulate
approximately 100 Earth years in 48 hours of wall-clock time using the same
number of cores. However, the speed of the radiation scheme decreases with
increasing number of bands and absorbers, and will therefore depend on the
set-up adopted for a particular planet.

In summary, SOCRATES gives us greatly improved flexibility to model atmospheres
with different composition and irradiation than present day Earth. However, due
to the desire to keep the computation time as small as possible while at the
same time calculating accurate fluxes and heating rates, some adaptation is
needed for each planet and star.

\subsection{Convection and Clouds}\label{subsec:Clouds}

The cumulus parameterization in \planet{} uses a mass flux approach that
requires both instability and a trigger based on the buoyancy of moist air
lifted a finite distance to initiate convection \citep[version ``AR5''
in][]{DelGenio2015}. In this sense it is more resistant to convection than
parameterizations in other planetary GCMs that require only instability to
initiate convection \citep[e.g.][]{Song2013}. The mass flux scheme utilizes a
cloud model that simulates the thermodynamic, dynamic and microphysical
properties of air rising in convective updrafts. The depth of convection is
determined by the distance the updraft penetrates above its level of neutral
buoyancy before the diagnosed convective updraft speed goes to zero. The
initial mass flux is calculated as that required to produce neutral buoyancy at
cloud base, with entrainment increasing the mass flux at higher levels and
detrainment decreasing the mass flux above the level of neutral buoyancy.
Simultaneous subsidence of the grid scale environment that adiabatically warms
and dries the gridbox to maintain subsaturated conditions, and convective
downdrafts formed from negatively buoyant mixtures of updraft and environmental
air, compensate the parameterized updraft mass flux. This approach differs from
adjustment schemes that seek to maintain a specified (sometimes saturated)
humidity profile and a moist adiabatic lapse rate. The differences in the mass
flux and adjustment approaches may affect estimates of the inner edge of the
habitable zone \citep{Wolf2015}.

Entrainment of subsaturated environmental air limits convection depth, but the
next generation of \rocke{} will include stronger entrainment that produces
more realistic subseasonal variability and cools and dries the tropopause
region relative to that in \planet{} \citep{DelGenio2016}. This may have
consequences for estimates of water loss for warm planets. Likewise the
\planet{} version transports condensed water upward too vigorously relative to
that which will be in the next generation of \modele, although this makes
little difference to reflected sunlight because these clouds are optically
thick \citep{Elsaesser2016}. For \rocke{} we have relaxed a limit in the parent
Earth \modele{} that restricts convection top pressures to $> \SI{50}{\hecto
\pascal}$.

Stratiform clouds in \planet{} have subgrid cloud fractions that are diagnosed
from local relative humidity and stability (\citealt{DelGenio1996} and updates
described in \citealt{Schmidt2006, Schmidt2014}). This differentiates our model
from planetary GCMs that require a gridbox to saturate before a cloud forms
that fills the gridbox. This is potentially important in estimates of the width
of the habitable zone, because the most important cloud feedback (and the one
that differs most widely among models) in terrestrial climate change
simulations is due to changes in cloud fraction \citep{Zelinka2016}. Cloud
water mixing ratios evolve prognostically based on simplified versions of
microphysical processes that are not easily generalized to treat cloud
processes on planets with different gravity or atmospheric pressure. The next
generation model will include a more explicit 2-moment microphysics
representation \citep{Gettelman2015} that can scale more easily to other
planets. 

\rocke{} is adjusted to planetary radiation balance using free cloud parameters
that regulate the onset of fractional cloudiness in the free troposphere and
boundary layer, and the rate at which small cloud liquid and ice particles are
converted to rain and snow that precipitate from the clouds. The specific
values of these tuning parameters for Earth are chosen to produce reasonably
accurate surface temperatures, but no such observational constraint yet exists
for exoplanets, and multiple choices that can bring the model to balance at
different temperatures are possible \citep[][see Figure 1]{Way2015}. Likewise,
more exotic planet configurations (e.g., synchronously rotating, zero
obliquity, etc.) may not come into balance at Earth free parameter settings and
are therefore adjusted as needed within reasonable ranges.

Only H$_{2}$O convection and clouds are represented in the baseline \rocke{}
model. For the next generation version we will allow for the possibility of
condensates such as CO$_2$ and CH$_4$ that are important on Mars and Titan, and
on exoplanets near the outer edge of the habitable zone (see Section
\ref{subsec:variableatmmass}).

\subsection{Planetary Boundary Layer}\label{subsec:PBL}

The planetary boundary layer (PBL) in \planet{} is described in
\citet{Schmidt2006}. It is based on nonlocal transport of dry-conserved
variables (virtual potential temperature and specific humidity). It includes a
diagnosis of the turbulent kinetic energy profile based on large-eddy
simulation studies and uses the resulting profile to define the PBL depth.
Cloud top sources of turbulence are not included, although the effect of
enhanced mixing at the top of cloudy boundary layers is estimated by the cloud
parameterization. The next generation \rocke{} will incorporate a full moist
turbulence scheme that transports liquid water potential temperature and total
water mixing ratio and includes cloud-top radiative cooling as a source of
turbulence. Boundary layer clouds have largely been absent from discussions of
exoplanet habitability to date, but considering that they are the largest
source of uncertainty in Earth’s climate sensitivity \citep{Zelinka2016}, they
warrant more attention in exoplanet studies.

\subsection{Cryosphere}\label{subsec:Cryosphere}

The cryosphere in \rocke{} –- encompassing areas of snow, land ice, and sea ice
-– has not been significantly modified from the modern Earth version of the GCM
\citep{Schmidt2006,Schmidt2014}, so hexagonal ice (ice Ih) is the only natural
phase of water ice represented. This means that \planet{} is not yet capable of
simulating the physical or spectral characteristics of water ices on worlds
with very low surface temperatures (below 75K; e.g., ice XI), and/or ice under
pressures of $\sim$200 MPa or more (e.g., ices II, III and IX)
\citep{Bartels2012}.

Snow may accumulate on any solid surface, including land ice and sea ice, as
long as surface conditions are sufficiently cold. Total snow column depth is
divided into two to three layers once the snow depth exceeds 0.15 m; as new
snow is added to the uppermost layer, older snow is redistributed to the
underlayers. Both heat and water are permitted to pass through the snow column
and into the ground (soil) beneath. Areas with snow depths $\leq$ 0.1 m are
considered to have only patchy snow cover. Also, snow cover over land depends
on local topography \citep{Roesch2001} and is never allowed to exceed 95\% of
the cell. If snow accumulates on top of either land or sea ice to a depth
greater than one meter, the bottom of the snow layer is compacted to ice. Note
that under cold global conditions snow mass may accumulate on land to such an
extent that the mass of the ocean is noticeably reduced; however, this ocean
mass reduction will not be detectable as a change in the global land/sea
fraction.

Land ice has the simplest treatment of the three cryosphere components in
\rocke. Where land ice is distributed as an initial input to the GCM, it
consists of two layers that may change thickness in response to mass balance
changes (accumulation minus sublimation and melting) induced by snow or rain.
However, \planet{} does not have dynamic land ice capabilities that would allow
the footprint area or the defined elevation of an ice sheet to grow or shrink
in response to forcings, or to affect ocean depth. A glacial melt
parameterization permits the return of land ice mass lost as meltwater to the
oceans, as well as calving of ``icebergs,'' into geographic-specific coastal
ocean cells. 

Sea ice extent and thickness may be prescribed for simulations with a specified
SST ocean model (Section \ref{subsubsec:specified-ocean}, or as an initial
condition for simulations with mixed-layer oceans (Section
\ref{subsubsec:qflux-ocean}) or dynamic oceans (Section
\ref{subsubsec:dynamic-ocean}). For the latter two types of simulations, it is
also allowed to develop as a consequence of other climate forcings on a world
that initially has no sea ice. When sea ice is allowed to respond to other
forcings, \planet{} uses the thermodynamic-dynamic sea ice formulation
described in \cite{Schmidt2006,Schmidt2014} to control its formation and
transport across the ocean surface. The formulation consists of four layers of
variable thickness but fixed fractional height, each of which has prognostic
mass, enthalpy, and salt content. Four sea ice surface types are included: bare
ice, dry snow on ice, wet snow on ice, and melt ponds \citep{Ebert1993}. Melt
pond mass accumulates as a fraction of surface runoff, decays on a time scale
that depends on the presence or absence of current melting, and re-freezes when
the temperature is below --10$\degr$C.  Unlike the modern Earth version of the
GCM, the sea ice thermodynamics in \planet{} do not yet include the effects of
internal brine pocket formation \citep{Bitz1999,Schmidt2014}, which would
improve energy conservation and result in thinner equilibrium ice thickness
compared to non-energy-conserving ice models \citep{Bitz1999}; this capability
will be introduced in a future release. 

Sea ice dynamics, especially important on synchronously rotating aquaplanets,
is treated with a viscous-plastic formulation for the ice rheology that takes
such factors as the Coriolis force, wind stress, ocean-ice stress, slope of the
ocean surface, and internal ice pressure into account when calculating strain
rates and viscosity for ice advection. Frazil ice (spicules or plates of new
ice suspended in water) is allowed to form either under existing ice or in open
ocean, as long as surface fluxes cool the water to the freezing point, given
the local salinity; once formed, the frazil ice advects along with the
previously existing sea ice. For a total sea ice thickness of five meters or
less, leads (narrow linear fractures) are allowed to form as a result of
shearing or divergent stresses. These leads can act as conduits for heat,
moisture and gas fluxes from the ocean below. 

As a planet's climate (and ultimately, its detectability via remote
observations) can be highly sensitive to the snow and ice albedo
parameterizations used in a GCM, we highlight here the key aspects of
\planet{}'s treatment of albedo in the cryosphere \citep[see
also][]{Schmidt2006}. 

The albedo of any surface type is dependent on the zenith angle across all
latitudes. The albedo of snow, whether it exists on bare soil, land ice, or sea
ice, is also a function of age; on sea ice, whether snow is wet (in the
presence of precipitation or melting) or dry, has an additional effect. Wet
snow and aging snow are both less reflective than dry or new snow, respectively
(see e.g. Table \ref{tab:table3}). The snow masking depth (i.e. the depth of
snow needed to completely counter the albedo properties of the underlying
surface type) depends on the underlying surface, though sea ice and land ice
are generally considered masked if covered by 0.1 m of snow.

The albedo values for sea ice are area-weighted averages for the different
surface types, resolved in six spectral intervals (Table \ref{tab:table3}).
Melt ponds, which significantly reduce sea ice albedo, are parameterized using
a pond fraction and depth that varies as a function of melt pond mass. Bare ice
albedo increases between assumed maximum and minimum values as the square root
of ice thickness. 

\begin{figure}
\centering
\includegraphics[width=0.8\textwidth]{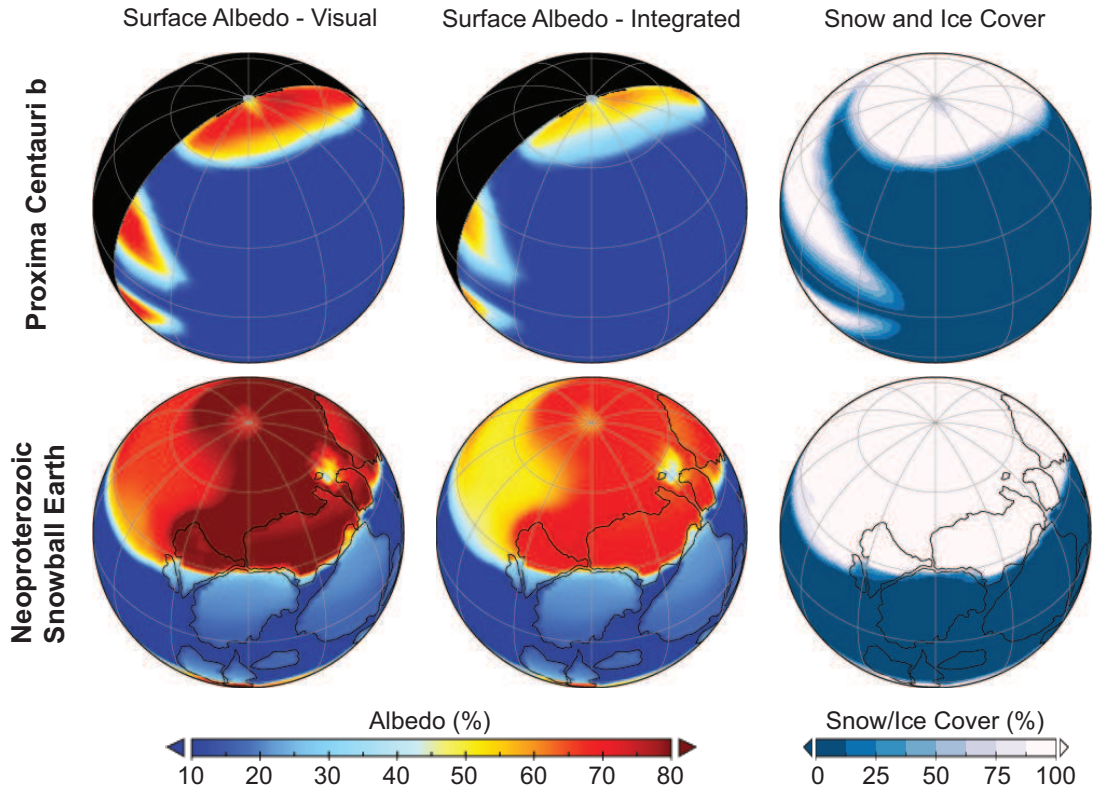}
\caption{Top row: Visible wavelength surface albedo map, spectrally integrated
surface albedo map, and snow and ice cover map for an aquaplanet simulation of
Proxima Centauri b using Planet 1.0 with the SOCRATES radiation scheme. Bottom
row: Corresponding maps for Neoproterozoic Snowball Earth. Snow cover over sea
ice or land produces the strongest surface albedo response in both
simulations.} \label{fig:Albedos}
\end{figure}

Although the surface albedo is resolved spectrally into the six bands, which
offers some sensitivity to different spectral irradiance, extinction of
radiation with depth through snow, ice, and liquid water currently only
distinguishes a “VIS” (290-690 nm) band and one “NIR” (690-1190 nm) band. The
extinction for each medium assumes solar-type surface irradiance fractions in
these bands and that radiation $>$1190 nm is not transmitted. Therefore, this
solar assumption will later be revised to capture sensitivity to alternative
stellar spectral irradiances.  Land ice, where it is not covered by snow, is
assumed to have a spectrally uniform albedo of 0.8. Land ice set as a boundary
condition for non-modern Earth simulations may use a different albedo value as
a default. 

\planet{} does not have the ability to change surface types (e.g., from ocean
to land or land to ocean). It therefore cannot treat situations in which sea
ice freezes to the ocean bottom along coastlines, or ocean mass decreases/sea
level falls as snow accumulates on land, situations which can occur at high
latitudes on low obliquity planets or for low instellations. To avoid this, for
\rocke{} experiments, ocean bathymetry is deepened along coastlines when needed
\citep[e.g.][]{Way2017}.

\begin{table}
\centering
\caption{Surface albedos of various sea ice surface types in different spectral intervals.}
\label{tab:table3}
\begin{tabular}{|l|l|l|l|l|l|l|}
\hline
Surface Type& VIS (nm) & NIR1 (nm) & NIR2 (nm) & NIR3 (nm) & NIR4 (nm) & NIR5 (nm) \\
\hline
            &330-770 & 770-860 & 860-1250 & 1250-1500 &
1500-2200 & 2200-4000\\
\hline \hline
Bare ice (min) & 0.05 & 0.05 & 0.05 & 0.050 & 0.05 & 0.03\\
\hline
Bare ice (max) & 0.62 & 0.42 & 0.30 & 0.120 & 0.05 & 0.03\\
\hline
Snow (wet)     & 0.85 & 0.75 & 0.50 & 0.175 & 0.03 & 0.01\\
\hline 
Snow (dry)     & 0.90 & 0.85 & 0.65 & 0.450 & 0.10 & 0.10\\
\hline 
Melt pond (min)& 0.10 & 0.05 & 0.05 & 0.050 & 0.05 & 0.03\\
\hline
\end{tabular}
\end{table}

\subsection{Chemistry}\label{subsec:Chemistry}

In simulating other planets, including early Earth, certain assumptions that
are built into \modele{} can become invalid, so updated or new
parameterizations need to be developed for \rocke{}. This is especially the
case for reduced atmospheres like those of Archean Earth, Titan and probably
Pluto. Currently \modele{} and \rocke{} are able to run with interactive gas
phase chemistry and a number of different aerosol configurations, from simple
bulk parameterizations to full aerosol microphysics calculations. Simulating
ice and gas giant atmospheric chemistry is beyond the capabilities and scope of
\rocke{}. The implementation of an automated solver of chemistry, which will
essentially allow the simulation of any atmospheric composition regardless of
its redox state, is under way. This involves the use of the kinetic
pre-processor \citep[KPP;][]{Sandu2006}, which will replace the scheme
described below, and is expected to gradually become available in \modele{} and
\rocke{} in coming years. Its adoption will enable the use of alternate
chemical schemes for Earth and planetary science applications, facilitating the
easy update and upgrade of the chemical mechanisms currently in the model.

The current chemical scheme in \modele{} only allows for calculations of
O$_2$-bearing atmospheres and is strongly linked with the expected composition
of the present-day atmosphere of Earth. The model uses the CBM-4 chemical
mechanism \citep{Gery1989}, which explicitly resolves the inorganic chemistry
that involves NO$_x$ and O$_3$, as well as the chemistry of methane and its
oxidation products \citep{Shindell2001}. In addition, it resolves the chemistry
of higher hydrocarbons via a highly parameterized scheme, based on CBM-4, with
only minor changes \citep{Shindell2003}, and that of halogens in the
stratosphere to account for the ozone hole \citep{Shindell2006}. The solution
of the chemical system is facilitated by the use of chemical families, which
assumes that dynamic equilibrium will be established among the species that are
very closely related and interchange extremely fast, like the O$_x$ family
species (O($^3$P), O($^1$D), O$_3$), the NO$_x$ family (NO and NO$_2$) and the
HO$_x$ family (OH and HO$_2$), which allows the accurate solution of the system
that includes species with lifetimes from sub-seconds to months or even years. 

The parameterizations of chemistry involve the solution of the chemical
kinetics equations, which is in principle independent of conditions. However,
some assumptions are made to make the solution of the partial differential
equations less stiff, which should not be violated in a different atmospheric
composition configuration. One of the most important assumptions is that
molecular oxygen is always in excess, so reactions that involve it happen
instantaneously. This is the case for the hydrogen radical and all alkyl
radicals in the model:

\ce{H + O2 -> HO2}

\ce{CH3 + O2 -> CH3O2}

\ce{R + O2 -> RO2}

Where R is any alkyl radical with more than one carbon atom. These reactions
are extremely fast \citep{Burkholder2015}, and the assumption that they
dominate other competitive loss processes of H, CH$_3$ and R is valid for
extremely low O$_2$ levels. For H, the competitive processes would be reactions
with O$_3$ or HO$_2$, both of which will go down with reduced levels or O$_2$,
while for the alkyl radicals the competitive processes would be reactions with
atomic O. O$_3$ would also decrease in a low-O$_2$ atmosphere. Even without the
assumption that the levels of the competitive oxidants will go down, the
reaction of O$_2$ is still the dominant loss of these radicals for O$_2$ levels
as low as 10$^{-6}$ or present atmospheric levels (PAL), based on their
reaction rates alone \citep{Burkholder2015}. 

\begin{figure}
\centering
\includegraphics[width=0.7\textwidth]{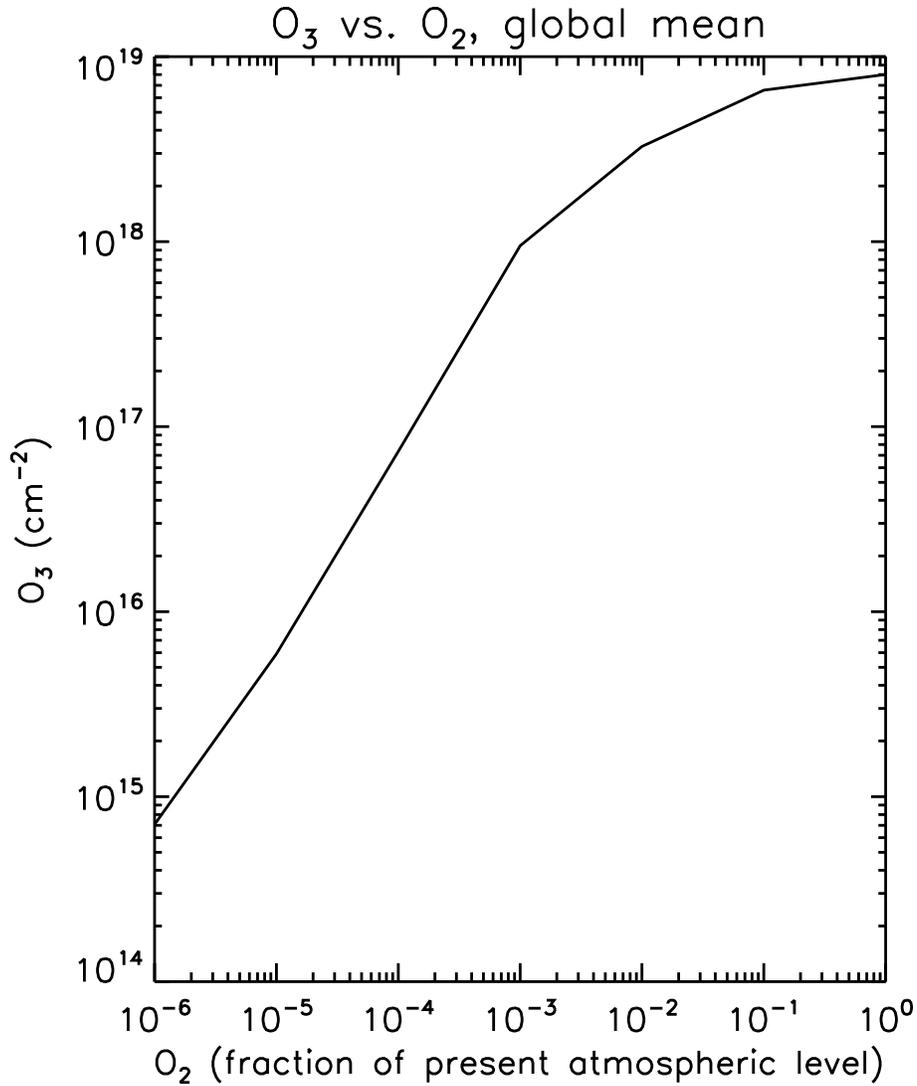}
\caption{Global mean O$_3$ column density as a function of O$_2$ concentration.}
\label{fig:O3summary}
\end{figure}

\begin{figure}
\centering
\includegraphics[width=0.7\textwidth]{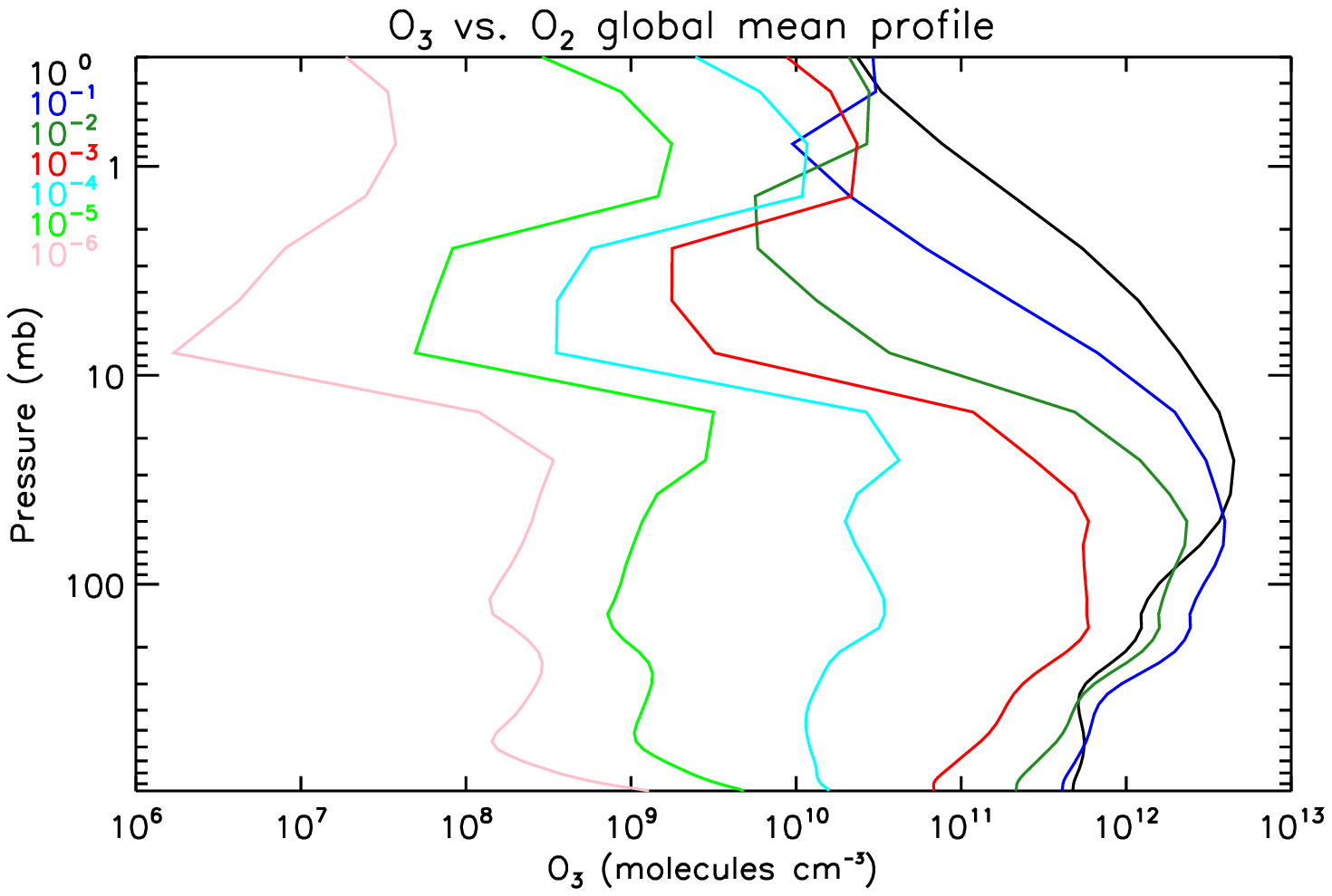}
\caption{Global mean O$_3$ column profile as a function of O$_2$ concentration.}
\label{fig:O3profile}
\end{figure}

We performed \rocke{} simulations of present-day Earth under pre-industrial
conditions (to minimize the impact of human activities on the atmospheric
state) in which we varied the levels of atmospheric O$_2$ from 1 to 10$^{-6}$
of PAL, to study how chemistry will be impacted, with a focus on O$_3$. We did
not allow radiation to be impacted directly by the O$_2$ changes, in order to
study the chemical response alone, but the effects of the results in O$_3$ were
included. The summary of the simulations is presented in
Fig.~\ref{fig:O3summary}, which agrees very well with the results of
\cite{KastingDonahue1979}. Interestingly, the calculated vertical profile of
O$_3$ for different O$_2$ levels (Fig.~\ref{fig:O3profile}) agrees with that of
\cite{KastingDonahue1979} only for O$_2$ levels down to 10$^{-3}$ PAL. The
model calculates a collapse of the stratosphere for O$_2$ levels below that
threshold because of the colder stratosphere that results from the decrease in
O$_3$, while the 1D model of \cite{KastingDonahue1979} does not. 

\subsection{Aerosols}

Simulating aerosols prognostically in other planetary configurations is also
possible with the GCM, with few modifications of the original scheme. For Earth
applications, the model contains carbonaceous (primary and secondary organic,
and black carbon) and non-carbonaceous (sulfate, ammonium, nitrate, sea salt,
and dust) aerosols. 

The carbonaceous aerosols can be formed either by direct emission in the
atmosphere or by the oxidation of precursor volatile organic compounds. In an
O$_2$-rich atmosphere, organic hazes like those of the Archean, Titan and Pluto
cannot be simulated in \planet{}. This is a limitation of the gas phase
chemistry of the model which cannot calculate the photochemical formation of
condensables in a reduced atmosphere, rather than a limitation of the aerosol
scheme. Methane is not able to form aerosols in oxidizing environments, so
unless there is life to form higher hydrocarbons (organic aerosol precursors)
or any combustible carbonaceous material which can inject organic and black
carbon particles in the atmosphere via burning, the carbonaceous aerosols are
not needed in non-Earth planetary configurations where O$_2$ is present. 

Non-carbonaceous particles are present on both Venus (sulfate aerosols, a
SO$_2$ oxidation product) and Mars (dust). Any planet with active volcanism is
expected to have some level of sulfate aerosols in their atmosphere, while any
planet with erodible bare rock is expected to have dust. In addition, any
planet with surface saline water and wave breaking is expected to have sea salt
aerosols injected into the atmosphere. 

During most of Earth's history there were salty oceans covering parts of the
planet, and the presence of an atmosphere ensures that waves would form, so the
presence of sea salt aerosols should be considered ubiquitous from very early
on. The GCM can interactively calculate sea salt aerosol sources in the
atmosphere using a variety of parameterizations \citep{Tsigaridis2013}; the
default \modele{} scheme is that of \cite{Gong2003} which is a function of
ocean salinity and surface wind speed over the oceanic grid cells. A
parameterization that takes into account sea surface temperature is also
available \citep{Jaegle2011}, but it is tuned towards present-day Earth
conditions, since there are no physical constraints on the process. Sea salt
aerosols are able to run as a standalone component in the model, without
requiring the presence of other aerosols, which could save significant
computational resources in simulating ocean worlds. 

As with sea salt, dust can be calculated interactively in \modele{}
\citep{Miller2006}. The source function depends on surface type and orography,
as well as wind speed. Topographic depressions tend to be good sources of dust,
contrary to mountain tops and steep slopes. For experimental unpublished
\rocke{} simulations of dust on Mars, we used
MOLA\footnote{http://pds-geosciences.wustl.edu/missions/mgs/mola.html}
topography data to construct a map of preferred dust sources based on the
topography of the planet, similar to what \cite{Ginoux2001} did for Earth. The
fraction of dust available for wind erosion implies that valleys and
depressions have accumulated dust, compared to flat basins where dust is more
homogeneously distributed. This is represented by calculating the probability
to have accumulated sediments, S$_i$, in a $1\degr\times1\degr$ resolution, as
a function of the minimum (z$_{min}$) and maximum (z$_{max}$) elevations of the
surrounding $10\degr\times10\degr$ topography of the grid box $i$, which has an
elevation of z$_i$:

\[ S_{i}=\Big(\frac{z_{max}-z_{i}}{z_{max}-z_{min}}\Big)^5 \]

\begin{figure}
\centering
\includegraphics[width=0.7\textwidth]{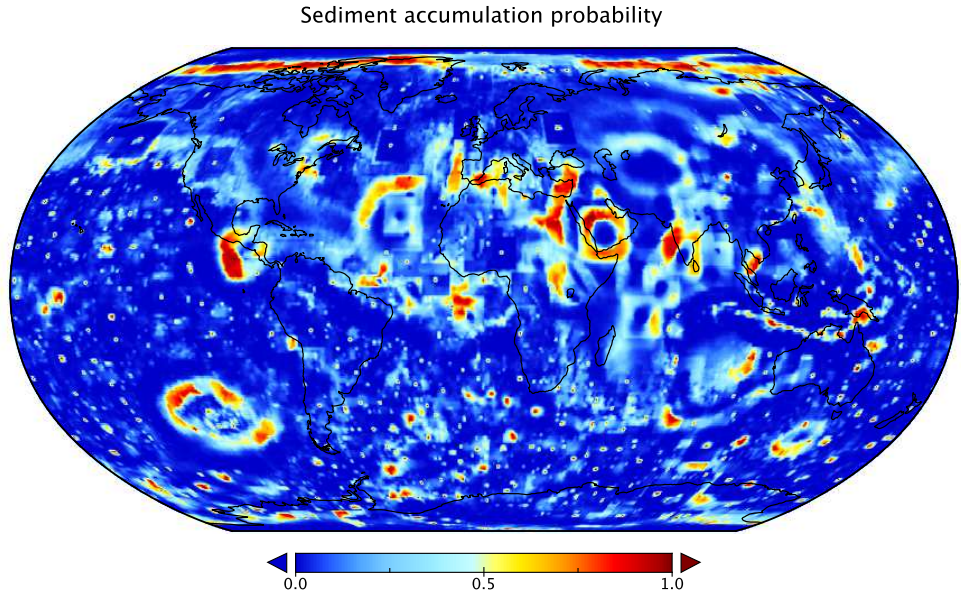}
\caption{Calculated accumulated sediment probability.}
\label{fig:mars_sediment}
\end{figure}

The calculated accumulated sediment probability is shown in Fig.~\ref{fig:mars_sediment}.

Active volcanism is another way to form aerosols in the atmosphere of the
planet. Volcanic eruptions inject large amounts of ash and SO$_2$ into the
atmosphere, among other constituents. Ash, which we are not yet able to
simulate in the model, is absorbing and the particles are usually large and
thus have too short a lifetime to be globally important, but SO$_2$ can form
sulfate particles in an oxidizing atmosphere, like those of Earth, Venus, or
Mars. Although the model is able to simulate both the formation of sulfate from
SO$_2$ and the lifetime of sulfate particles in the atmosphere, the sources of
SO$_2$ from active volcanism on other planets are virtually unknown, making its
interactive simulation difficult. The sulfur cycle on Mars will be studied in
the future.

\section{Geophysical properties} \label{sec:properties}

\subsection{Land/Ocean Distribution and Topographic Relief}

The land/ocean mask used by the GCM can be defined using a fractional or
non-fractional method, with the former preferred at coarse resolutions when
shorelines are irregular and where ocean gateways may have significant climate
impacts. A fractional land mask scheme signifies that an individual grid cell
can be defined not only as \SI{100}{\percent} land or \SI{100}{\percent} ocean,
but also as some percentage of each. This means that coastal grid cells are
treated by other routines in the model as some portion ocean and some portion
land. Topographic relief then is a weighted average based on the elevation of
the land fraction and zero for the ocean fraction of the cell. Available input
data sets for paleoclimate and other planets are described in Appendix
\ref{subsec:GCMInputs}, and users may create their own.

\subsection{Continental Drainage (Runoff)}

Riverflow and continental drainage redistributes freshwater via the water cycle
and thus impacts soil moisture, which affects land temperatures and
precipitation, and the ocean salinity distribution, which impacts ocean
circulation. For simulations that use modern Earth land/ocean distributions, we
use the same riverflow and drainage patterns as defined for modern Earth
climate experiments. Drainage directions in the GCM for non-modern-Earth
continental configurations must be assigned via a custom input file. We
generate the new drainage patterns by examining the topographic elevation
boundary condition array and, working inward from continental edges, we
calculate the slope of each continental grid cell in eight directions (four
sides, four corners). Runoff is then removed from each cell in the direction of
maximum slope, tracing a route back to the coast. For coastal grid cells that
have more than one border adjacent to an ocean grid cell, runoff crosses the
coastal grid cell on the same trajectory as in the adjacent inland grid cell.

Lakes may be treated as pre-defined static features, or be permitted to grow
and shrink dynamically in response to rainfall. In dynamic mode, lakes may also
develop in topographic lows, with drainage developing once the water level
rises above the lowest edge of the lake basin. Where a drainage route does not
already exist, excess lake water runs off by means of the local drainage
patterns defined above.

In \modele, glacial ice melts directly from the Greenland and East Antarctic
ice sheets, and enters the surface ocean in prescribed cells wherever the edges
of the ice sheets coincide with continental edges. However, this can be
adjusted for a given topography or other needs by specifying the geographic
locations where freshwater (from ice melt) is distributed back into the oceans.

\subsection{Ground hydrology, albedo, and surface vegetation}

Ground hydrology employs a 6-layer soil heat and water balance scheme
\citep{abramopoulos1988,RosenzweigAbrampoulos1997}, and the approach for
calculating underground runoff is described in \citet{aleinov2006}. The surface
energy and water balance algorithm calculates heat and water  content on an
explicit numerical scheme in the soil layers and vegetation canopies (if
present) to predict temperatures and saturation. In current implementation the
total soil depth is 3.5~m with the boundary at the bottom impermeable to both
heat and water. Surface spectral albedo is partitioned currently into the same
6 broad bands shown in Table \ref{tab:table3}: (\SIrange{300}{770}{\nano
\meter}, \SIrange{770}{860}{\nano \meter}, \SIrange{860}{1250}{\nano \meter},
\SIrange{1250}{1500}{\nano \meter}, \SIrange{1500}{2200}{\nano \meter}, and
\SIrange{2200}{4000}{\nano \meter}), in an area-weighted average for cover
types including vegetation, bare soil (with regard to albedo, see below) and
permanent ice. For questions regarding extrasolar planets, the ground hydrology
boundary conditions that must be modified include soil texture and albedo maps
for bare soil on a lifeless planet, or albedo influenced by vegetation. 

The land albedo is spectrally resolved in the same VIS and NIR  bands as
described in Table \ref{tab:table3}.  The albedo can be calculated in 2
different ways, which requires different sets of input files:
\begin{enumerate}
\item The Lambertian albedo scheme from \citet{Matthews1984}, for which an
input file gives grid fractional areas of land cover types, including
vegetation types and bare soil. Bare dry soil albedo is specified as a
combination of ``bright" and ``dark" soil of albedo 0.5 and 0.0, respectively,
so that their area-weighted averages gives the soil albedo. Soil albedo is
spectrally flat and is assumed to depend linearly on soil saturation, becoming
twice lower for a completely saturated soil. This is the scheme used since 1984
through to \citet{Schmidt2014}. These input files are suitable for Earth
vegetation and the Solar spectrum. If simulating Earth vegetation under other
stars, users should revise the broadband albedos to account for different band
irradiances from different stellar types. 
\item A zenith angle-dependent surface albedo scheme described in \citet{NiMeister2010881}. If vegetation
cover is prescribed, then maps of vegetation cover, vegetation height, maximum
leaf area index, and soil albedo are separate input files. When ecological
dynamics are turned on  vegetation cover does not need to be initialized, since
the vegetation will grow and die according to climate interactions. The soil
albedo map allows resolution of soil into the 6 spectral bands of Table
\ref{tab:table3}. End member broad band optical properties should be calculated
to take into account different stellar spectral types, as described in Appendix
\ref{subsec:StellarSpectra}.
\end{enumerate}

Surface life, particularly photosynthetic life, can strongly influence a
planet's surface properties like spectral albedo, as we know from the
vegetation red-edge on Earth \citep{Tucker1976,Kiang2007a}. The Ent Terrestrial
Biosphere (Ent TBM) is the Earth dynamic global vegetation model (DGVM)
currently coupled to \modele{} \citep{Schmidt2014,Kim2015}. While it can be an
interesting exercise to subject Earth vegetation\footnote{The Ent TBM can allow
any number of user-defined plant functional types, and supports parameter sets
for 13 Earth types.} with full ecological dynamics to conditions on another
planet to see what survives, seasonality and physiological differences between
plant types are based on close adaptations to the star-planet orbital
configuration and climatic regimes, so it would be inappropriate to utilize the
current Sun-Earth based plant functional types (PFTs) for extrasolar planets. 

As part of proposed work for \rocke, an Exoplanet Plant Functional Type
(ExoPFT) is being introduced to provide a ``generic plant'' for simulations of
alien vegetation influences on exoplanets. This ExoPFT will simply ``find the
water'', i.e. provide surface life wherever the planet is habitable. This
generic plant will be similar to C3 annual grasses currently in the Ent TBM,
but will have easily modifiable physiological and optical properties to allow
experimentation with the potential distribution of life over a planet's land
surfaces, its impact on the surface energy balance and surface conductance, and
its possible detectability. The ExoPFT will be a simple, non-woody, vascular
plant with roots to access soil water that simulates the very basic influences
of vegetation on climate: surface albedo and water vapor conductance. To ``find
the water", the ExoPFT will be parameterized to emerge and senesce according to
the presence of water, with broad climatological tolerance, and user-specified
leaf spectral albedo to investigate effects on the climate of photosynthetic
pigments adapted to alternative parent star spectral irradiance (e.g.,
adaptation behavior similar to that proposed by \citet{Tinetti2006} and
\citet{Kiang2007b}. This ExoPFT will be built within the platform of the Ent
TBM.

The Ent TBM currently provides the vegetation biophysics and land carbon
dynamics to \modele{} \citep{Schmidt2006,Schmidt2014}. The ExoPFT will utilize
the EntTBM scheme for vegetation conductance of water vapor and CO$_2$, and
leverage a new canopy radiative transfer model being added to the Ent TBM. In
addition, the ExoPFT’s phenology (timing of leaf-out and senescence) and growth
scheme will introduce its water-seeking parameterization within the Ent TBM
framework.

Plant photosynthesis is sensitive to the atmospheric CO$_2$ surface
concentration. Leaf conductance of water vapor, which is coupled with
photosynthesis, is inversely proportional to surface CO$_2$ concentrations.
These sensitivities are represented in the Ent TBM biophysics via the
well-accepted \citet{Farquhar1982} photosynthesis model coupled with
\citet{Ball1987} leaf stomatal conductance detailed in \citet{Kim2015}. This
inverse relation to CO$_2$ is infeasible for an atmosphere with zero CO$_2$,
which would not be realistic for a planet with photosynthesis. Numerically in
the GCM the lowest CO$_2$ level recommended is 10 ppm. This is the CO$_2$
compensation point where photosynthesis and respiration just balance each
other. This is typical for C4 photosynthesis, a type of photosynthetic carbon
fixation pathway that enables uptake of CO$_2$ at lower atmospheric
concentrations than the other common pathway known as C3 photosynthesis.
Coupling to the atmosphere currently relies on roughness lengths determined by
the ground hydrology scheme for the GCM grid cell scale. 

Scaling leaf conductance as well as optical properties to the canopy scale for
the ExoPFT will be done with the new prognostic vegetation canopy radiative
transfer scheme, the Analytical Clumped Two-Stream (ACTS) model
\citep{NiMeister2010881,Yang2010895}. This model has recently been incorporated
in the Ent TBM. The ACTS model depends on zenith angle, direct/diffuse
partitioning of radiation, canopy structure,\footnote{The canopy structure
includes time variation in leaf area index, canopy height stratification, and
plant densities.} and end member spectral optical properties of foliage, soil,
and snow. The prior canopy radiative transfer scheme described in
\citet{Schmidt2006,Schmidt2014} relies on prescribed seasonal canopy albedos by
vegetation type with fixed seasonal Leaf Area Index (LAI) \citep{Matthews1984}
and is not a function of dynamic LAI, and therefore is not suitable for use
with dynamically changing vegetation.

End member optical properties are summarized into the same 6 broad bands used
for the ground hydrology (see Table \ref{tab:table3}). Alteration of these band
albedos must take into account the stellar spectral irradiance, particularly if
otherwise investigating the same vegetation optical properties but with
different parent stars. For example, the ACTS Earth vegetation end member
broadband spectra (leaf reflectance and transmittance) are derived from
convolving hyperspectral leaf data with a solar surface irradiance spectrum at
60 degrees zenith angle (approximating an average over the illuminated face of
the planet) with a U.S. standard atmosphere.

Ent TBM vegetation dynamics of phenology (seasonality) and growth have been
tested at the site level for several Earth plant functional types
\citep{Kim2015}. The ExoPFT phenology will be parameterized simply to leaf out
and senesce with the availability of water, without other mortality and
establishment drivers than water (i.e. insensitive to the plant’s carbon
reserves, since this balance is already poorly known for Earth plants).
Ecological dynamics involving competition, fire disturbance, and establishment
will not be necessary to drive vegetation cover change, since only one ExoPFT
will represent vegetation, driven by water availability.

\subsection{Variable Atmospheric Mass}\label{subsec:variableatmmass}

Typically, the atmosphere contains one or more components that may
condense/evaporate at the surface of the planet or within the atmosphere. One
can distinguish three cases: (1) A dilute (small fraction of total atmospheric
mass) condensable gas. This is the case for water vapor on modern Earth. (2) A
single-component atmosphere that consists of a gas that condenses at typical
temperatures and pressures. (3) A non-dilute (significant fraction of total
atmospheric mass) condensable gas. In the first case changes in atmospheric
mass and heat capacity due to condensation/evaporation can be neglected except
in the cumulus parameterization, where the effects of water vapor and
precipitation loading on parcel buoyancy are non-negligible. The processes at
the surface in this case will typically be governed by turbulent diffusion
fluxes. 

Modern Mars, where CO$_2$ accounts for most but not all of the atmospheric
mass, is actually an example of case 3.  For \planet{} we have taken the first
steps toward creating a Mars GCM by ignoring the minor constituents and
treating Mars as a pure CO$_2$ atmosphere, corresponding to case 2. In this
case the change in the atmospheric mass over the seasonal cycle can be
significant and cannot be neglected for calculating temperatures and pressure
gradients. Also, the amount of condensable substance at the surface is
abundant, so the process of condensation/sublimation is governed by the energy
balance, rather than by the diffusion fluxes. In the remainder of this section
we present the algorithm we use to model the condensation of a condensable
single-component atmosphere at the planet's surface. The description of similar
processes for a dilute condensable component in \modele{} can be found in
\citet{Schmidt2014}.

We assume that the condensate is stored in the upper soil layer(recall that
\modele{} has 6 soil layers). We also assume that the formation of the
condensate is controlled purely by energy balance and the matter is condensed
or sublimated as needed to compensate for energy loss or gain by the upper soil
layer. Once formed the condensate is assumed to stay at the condensation
temperature $T_\textrm{cond}$, which depends on the atmospheric pressure $p_s$
at the planet's surface. This temperature is described by the
Clausius-Clapeyron relation. For the case of CO$_2$ condensation on Mars it can
be expressed approximately via \citet{Haberle1982}:

\begin{equation}
  T_{\textrm{cond}}(p_s) = 149.2 + 6.48 \ln(0.135 \ p_s)
  \label{eq:T_cond}
\end{equation}
where $T_{\textrm{cond}}$ is in Kelvin and $p_s$ is in millibars.
We define the latent heat of condensation $L_c$ as the amount of heat
needed to melt a unit mass of condensate and bring it to surface air temperature $T_s$
\begin{equation}
  L_c(T_s,p_s) = L_{c0} + c_{pg} (T_s - T_0) - c_{pc} (T_{\textrm{cond}}(p_s) - T_0)
  \label{eq:L_c}
\end{equation}
where $c_{pg}$, $c_{pc}$ are the specific heat capacities of the condensable
substance in gaseous and condensed form respectively. $L_{c0}$ is the latent
heat of condensation at some fixed temperature $T_0$.  For CO$_2$ condensation
on Mars we use:
\begin{align*}
  c_{pg} &= \SI{770.2}{\joule \per \kilo \gram \per \kelvin} \quad \text{\citep{Lange10}} \\ 
  c_{pc} &= \SI{1070.7}{\joule \per \kilo \gram \per \kelvin} \quad \text{\citep{Giauque1937}} \\
  L_{c0} &= \SI{5.902e5}{\joule \per \kilo \gram} \quad  \text{\citep{Haberle1982}} \\
  T_0 &= \SI{150.0}{\kelvin}
\end{align*}

The prognostic variable which defines the ground temperature and the amount of
condensate stored in the first layer of soil is the amount of energy per unit
area in this soil layer $H_1$. The quantity $H_1$ is defined with respect to
some reference temperature $T_{\textrm{ref}}$, in a sense that the substance at
the temperature $T_{\textrm{ref}}$ has energy zero. In our model we set
$T_{\textrm{ref}} = 273.15$, since it helps in dealing with freezing/thawing of
water in Earth simulations using \modele{}, but one can choose any reference
temperature above the condensation point. The ground temperature $T_g$ can be
obtained as
\begin{equation}
  T_g = \max \left( \frac{H_1}{c_{\textrm{soil}} \Delta z_1} + T_{\textrm{ref}},\ T_{\textrm{cond}}(p_s) \right)
  \label{eq:T_g}
\end{equation}
where $c_{\textrm{soil}}$ is the volumetric specific heat capacity of soil and $\Delta  z_1$ is the thickness of the upper soil layer. If
\begin{equation}
  \frac{H_1}{c_{\textrm{soil}} \Delta z_1} + T_{\textrm{ref}} < T_{\textrm{cond}}(p_s)
\label{eq:H}
\end{equation}
then a non-zero amount of condensate is present, and its mass per unit area can be computed as
\begin{equation}
  m_{\textrm{cond}} = - \frac
  {H_1 - c_{\textrm{soil}} \Delta z_1 (T_{\textrm{cond}}(p_s) - T_{\textrm{ref}})}
  {L_c(T_s,p_s) -c_{pg} (T_{\textrm{cond}}(p_s) - T_{\textrm{ref}})}
  \label{eq:m_cond}
\end{equation}
Since we are dealing with a non-dilute case, the atmospheric pressure is affected by the formation of the condensate, which can be expressed as
\begin{equation}
  \frac{d p_s}{d t} = - g \frac{d m_{\textrm{cond}}}{d t}
  \label{eq:dp_s_dt}
\end{equation}
where $g$ is the gravitational acceleration. The heat content $H_1$ is controlled by the energy balance at the surface
\begin{equation}
  \frac{d H_1}{d t} = R_n - S - G + \frac{d m_{\textrm{cond}}}{d t} c_{pg} (T_s - T_{\textrm{ref}})
  \label{eq:dH_1_dt}
\end{equation}
where $R_n$ is net absorbed radiation at the surface, $S$ is the sensible heat
flux to the atmosphere, $G$ is the ground heat flux to the lower soil layers
and the last term on the right-hand side is the energy flux due to the
gain/loss of the substance from/to the atmosphere (which is assumed to be at
temperature $T_s$).

The algorithm described above is implemented as follows. At each time step
$H_1$ is first updated according to Eq.~(\ref{eq:dH_1_dt}) with the assumption
that there is no change in the amount of  condensate and the condition in
Eq.~(\ref{eq:H}) is checked. If true, the system of equations
Eqs.~(\ref{eq:T_cond}) to (\ref{eq:dH_1_dt}) is solved iteratively to obtain
the new values for $m_\textrm{cond}$, $T_g$, $p_s$. The change in the
condensate $\Delta m_\textrm{cond}$ over the time step is then computed as
\begin{equation}
  \Delta m_\textrm{cond} = m_\textrm{cond} - m_{\textrm{cond},0}
  \label{eq:delta_m_cond}
\end{equation}
where $m_{\textrm{cond},0}$ is the amount of condensate at the end of the
previous time step. $\Delta m_\textrm{cond}$ is then subtracted from the mass
of the condensable component of the lower atmosphere layer. The removed/added
gas is assumed to be at $T_s$, so the temperature of the lower layer of the
atmosphere is updated accordingly.  If condition (\ref{eq:H}) is false then no
condensate is present. If condensate was present at the previous time step,
then the corresponding amount of gas should be added to the lower atmospheric
layer and its temperature should be adjusted accordingly.

\begin{figure}
    \centering
    \includegraphics[width=0.6\textwidth]{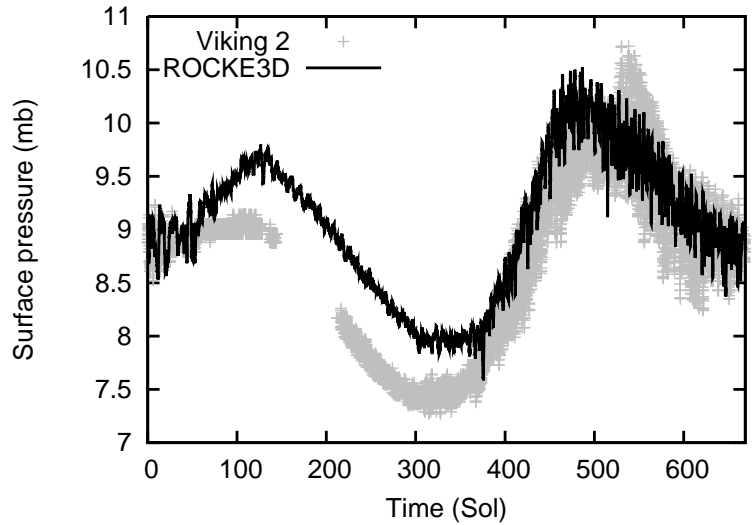}
    \caption{Annual cycle of Mars surface pressure as measured by the Viking 2
lander (gray crosses) \citep{Hess1977,Tillman1989} and surface pressure
simulated by \rocke{} (black solid line).
}\label{fig:MarsPressure}
\end{figure}

Figure~\ref{fig:MarsPressure} shows the annual cycle of surface pressure on
Mars at the location of the Viking 2 lander and that simulated with the surface
condensation routine activated in a version of \planet{} that includes some but
not all of the physics that affects Mars' climate (i.e. it uses the GISS
radiation scheme, which has limitations in treating atmospheres with
composition very different from Earth, it does not yet allow for CO$_2$ clouds,
and it does not yet incorporate dust).  Despite these limitations, the timing
and amplitude of the seasonal variation \citep{SharmanRyan1980} are in
reasonable qualitative agreement with observations. The site of the Viking 2
lander was chosen for comparison, because it is largely a flat area and can be
well represented by a coarse-resolution GCM cell such as that used in this
simulation. Most of the other landing sites have a more complicated terrain and
would require higher horizontal resolution for such  simulations, which is
beyond the scope of our current experiments.

In the description of the algorithm above (for simplicity's sake) we assume
that only one (non-dilute) condensable component is present, but our model can
also handle the presence of another (dilute) component. This is done by
including the latent heat due to the dilute component in equations
(\ref{eq:T_g}) to (\ref{eq:m_cond}) and including the dilute condensate heat
capacity into $c_{\textrm{soil}}$. Otherwise the dilute component itself is
treated as in \modele. The presence of both such components is necessary for a
more representative Mars simulation where both dilute (water) and non-dilute
(CO$_2$) condensable components are present. We note that Mars' atmosphere also
contains several non-condensing minor constituents, e.g., N$_2$. These are not
important for the dynamics, but do affect the ability of CO$_2$ to
supersaturate and thus the occurrence of CO$_2$ cirrus clouds
\citep{Colaprete2008}. This capability does not yet exist but will be added in
future generations of \rocke.

Currently, \rocke{} does not have the capability to treat case 3, i.e.
condensable constituents that represent a significant and variable fraction of
the mass of a multi-component atmosphere. This can become important as an
Earth-like planet approaches the inner edge of the habitable zone where
H$_{2}$O becomes a non-negligible part of the total atmospheric mass. This will
in turn affect pressure gradients and the thermodynamic properties of air and
will introduce non-ideal gas behavior. This feature will be added in a future
generation of \rocke. 

\section{Enhancement to Earth System Modeling (\modele) as a result of \rocke}
\label{sec:enhancement}

Generalizations and extensions to \modele{} to accommodate the requirements of
non-Earth planets can also benefit the Earth model through accelerated
implementation of previously planned user-facing improvements to flexibility
and accuracy.  With a view to future development work and its multi-planet
scope, this process also provides an opportunity for restructurings that
enhance programmer-facing ``code quality.''   A visible example of all these
trends is the reorganization of the time-management system, discussed in
Section \ref{sec:calendar}.

Other examples can be found in the modularization of the manner in which the
features of a planet are specified by a user.  \modele{} had previously required
the presence of input files associated with all surface types (which is
inconvenient for desert worlds and aquaplanets) and the time-space distribution
of radiatively active constituents important for Earth but not for other
planets (e.g. O$_{3}$).  Ongoing effort to increase the flexibility in the
specification of inputs and boundary conditions for Earth runs was extended to
cover additional use cases.

Improvements to \modele{} accuracy can sometimes result from running its modules
under conditions sufficiently different than those for Earth to expose
inappropriate approximations and/or coding errors.   In the first category, the
performance of the GISS Long Wave radiation scheme under conditions of extremely low column
water vapor (e.g. the Arctic and Antarctic) was improved via better look-up
tables generated in response to reports of problems in a cold and dry non-Earth
simulation.  In the second category, an aquaplanet simulation revealed some
oversights in the ocean horizontal diffusion of momentum.

Looking forward, an example of development planned for the Earth model that is
also highly convenient for non-Earth simulation includes the option for dynamic
surface-type masks due to factors including: sea level change, sea ice which
has thickened to the ocean bottom, and expansion/retreat of glacial ice.
While ``transient'' simulations of exoplanets in response to imposed time-varying
forcings are not a likely near-term objective, and the trajectory followed by a
model as it approaches equilibrium for a given set of imposed forcings is
typically not of interest either, it is convenient to have a model find its
equilibrium in a fully automated manner.   A brute-force procedure requiring
the user to try a sequence of prescribed land/sea distributions and associated
inputs greatly slows the rate at which equilibria can be determined.

Another advance that will benefit the Earth model is the use of the kinetic
pre-processor \citep[KPP;][]{Sandu2006} for interactive chemistry calculations
in \rocke{}. Its adoption will enable the use of alternate chemical schemes for
both Earth and planetary applications, facilitating the easy update and upgrade
of the chemical mechanisms currently included in the model.

\section{Appropriate use of \rocke} \label{sec:use}

\paragraph{Time scale:} As this is a GCM that simulates dynamics at time steps
of \SI{450}{\second} and parameterized physics at time steps of
\SI{30}{\minute} (and less in some submodules), it is best used for
scientific questions investigating time slice equilibrium climate behavior at
the scales of decades to centuries. The equilibrium time needed for ocean dynamics
can take much longer (some simulations require thousands of years), but the climate
characteristics are generally summarized over the last few decades of the run.
In some rare instances simulations tracking secular changes over 1000s of years
can be accommodated with this GCM (see \citealt{Way2017} for examples).
Geological time scale phenomena over millions of years, such as the changes in
the carbonate-silicate cycle, cannot be simulated by a GCM, but time slice
atmospheric composition conditions or flux rates could be prescribed.

\paragraph{Atmospheric escape:} The \rocke{} model top is at \SI{0.1}{\hecto
\pascal} ($\sim \SI{65}{\kilo \meter}$ for Earth), with 17 layers in the 40
layer model above the tropopause cold trap for Earth-like planets. This is
sufficient to resolve the stratospheric general circulation, which becomes
important for planets orbiting M stars in which significant shortwave
absorption by water vapor occurs at high altitude \citep{Fujii2017}. This
altitude is however tens of kilometers below the homopause, where
photodissociation of species such as H$_2$O and O$_2$ becomes important. Thus
\rocke{} cannot directly simulate atmospheric escape processes; this would
require coupling to upper atmospheric models specifically intended to simulate
ionization and escape processes \citep[e.g.][]{Gronoff2011}. Furthermore, since
\rocke{} (like all GCMs) can only simulate time slices of hundreds to thousands
of years, it cannot be used directly to address problems of atmospheric
evolution such as water loss in moist greenhouse states near the inner edge of
the habitable zone. Instead, GCM stratospheric water vapor mixing ratios are
traditionally compared to the threshold first estimated for 1-dimensional
models by \citet{Kasting1993} to characterize planets that may be at risk of
significant water loss \citep[e.g.][]{Kopparapu2016}. However, more
sophisticated approaches \citep[e.g.][]{Wordsworth2013} may be feasible.

\section{Discussion} \label{sec:discussion}

The use of GCMs to study the climate and weather of other planets has increased
dramatically in the past few years in response to increased interest in the
past climates of terrestrial Solar System planets, the rapidly growing list of
rocky and potentially habitable exoplanets, and the promise of more
discoveries, as well as atmospheric characterization of exoplanets by upcoming
and planned future spacecraft missions.  Every GCM has specific strengths and
weaknesses in its ability to simulate other planets and limitations in the
range of problems to which it can be applied.  The Earth climate modeling
community has found that as a result, a diverse population of GCMs offers
advantages over any single model by revealing robust behaviors that are common
to all models and appear to be determined by fundamental well-understood
physics, as well as features that differ among models due to differing
assumptions in the parameterized physics that highlight more poorly understood
processes.

The advantages of \rocke{} relative to other planetary GCMs are that its
physics is identical to the most recent published version of its parent Earth
GCM, it will remain current with future generations of the Earth model, and its
developers include a subset of the people who develop the Earth model. Thus it
includes much in the way of recent thinking about climate processes that
operate to determine Earth's changing climate, and its coding structure has
been generalized to easily allow simulations in parameter settings appropriate
to other planets without sacrificing process understanding.  It will also be
the first exoplanet GCM to represent basic functions of plants that should be
generally applicable to any habitable planet (for mock observations based on
GCM output, see Appendix \ref{sec:post-processing}). 

That having been said, \rocke{}'s Earth heritage produces limitations on its
use as well.  Some of these are structural and cannot easily be modified. The
most obvious is that \rocke{} is based on a model that is designed to simulate
only shallow atmospheres and oceans (i.e., much thinner than the planet radius)
with equations of state appropriate to such fluids.  Thus, \rocke{} can be
applied to planet sizes up to the super-Earth range, though not to
``waterworld'' planets on which water is a significant fraction of the planet
mass and a transition from water to ice at high pressure occurs.  Likewise, it
cannot be used to simulate or predict the transition from super-Earths to
sub-Neptunes with thick H$_2$ envelopes, nor can it simulate giant exoplanets.

Other limitations are specific to the \planet{} version of \rocke{} and will
disappear as future generations of the model are developed.  \planet{} has been
applied thus far only to planets with atmospheres composed of constituents
found on Earth at pressures equal to or less than that of Earth's atmosphere
and temperatures not too far from those present during Earth's history, such as
snowball Earth periods \citep{Sohl2015} and a hypothetical habitable ancient
Venus \citep{Way2016}. With the SOCRATES radiation scheme, it is now
sufficiently general to handle non-oxygenated atmospheres with prescribed
elevated greenhouse gas concentrations such as Archean Earth, and Earth-like
planets orbiting M-stars \citep{Fujii2017,DelGenio2017}. It has also been run
under variable eccentricity \citep{Way2017} and rotation periods as slow as 256
days as well as synchronous rotation, and a baseline modern Mars model has also
been created.  Rotation periods less than Earth's are also possible, but
require the higher horizontal resolution version of the model to accurately
capture the dynamics. Development under way will give it the capability to
simulate dense CO$_2$ atmospheres.  In its current form the model cannot
simulate atmospheres near the inner edge of the habitable zone, both because
the radiation does not include information from high-temperature line lists and
because the model does not treat the effects on atmospheric mass,
thermodynamics and dynamics of water vapor concentrations that are a
non-negligible fraction of the total atmospheric mass. Atmospheres with
compositions fundamentally different from those mentioned above (e.g., H$_2$-
dominated) are not yet available, although this is only a matter of developing
appropriate radiation tables for such planets.  Yet even in its current form
\rocke{} is well suited to address a wide range of science questions about
habitable and inhabited planets and should be a valuable tool for interpreting
near-future spacecraft observations of planets both within and outside the
Solar System and for supporting the planning of a possible future direct
imaging exoplanet mission.

\appendix

\section{GCM Inputs} \label{subsec:GCMInputs}

Users should consult the GISS GCM Software page for user guides on \modele,
necessary input data sets, and how to run the
model.\footnote{http://www.giss.nasa.gov/tools/modelE} For the Planet 1.0
branch, we offer the following data sets, tools and guidance for simulating
other planets.

\subsection{Surface pressure and gas amounts}

The surface pressure and amounts of atmospheric constituent gases are provided
as input to the model in the form of the total surface pressure,
$P_\text{tot}^\text{surf}$, and number/volume gas mixing ratios, $r_i$, for
each species $i$, respectively. The mean molecular weight of dry air can then
be calculated as
\begin{equation}
\bar m = \sum_{i=1}^{N_\text{s}} r_i m_i ,
\label{eq:mmw}
\end{equation}
where $N_\text{s}$ is the number of atmospheric constituent species, and $m_i$
is the molar weight of each gas. As the gases are assumed to be ideal, the
number mixing ratios are directly related to the partial pressures
$P_{\text{p},i}^\text{surf}$ of each gas at the surface:
\begin{equation}
P_{\text{p},i}^\text{surf} = r_i P_\text{tot}^\text{surf}.
\label{eq:pp}
\end{equation}
To calculate the contribution of each gas to the total surface pressure,
however, the molar weight of each gas needs to be taken into account. From
hydrostatic equilibrium the total surface pressure is given by
\begin{equation}
P_\text{tot}^\text{surf} = Mg/A = \frac{g}{A} \sum_{i=1}^{N_\text{s}} M_i,
\end{equation}
where $M$ is the total mass of the atmosphere, $A$ is the surface area of the
planet, $g$ is the acceleration of gravity and $M_i$ is the total mass of each
constituent gas. For well-mixed gases $M_i = \zeta_i M$, where $\zeta_i$ is the
mass mixing ratio of species $i$ and is related to the number/volume mixing
ratio by $\zeta_i = r_i m_i/\bar m$. Consequently, we can write the total
surface pressure as
\begin{equation}
P_\text{tot}^\text{surf} = \frac{g}{A} \sum_{i=1}^{N_\text{s}} \zeta_i M
= \frac{M g}{A} \sum_{i=1}^{N_\text{s}} r_i m_i/\bar m
= P_\text{tot}^\text{surf} \sum_{i=1}^{N_\text{s}} r_i m_i/\bar m,
\end{equation}
and the contribution to the surface pressure of gas $i$ is therefore given by
\begin{equation}
P_i^\text{surf} = P_\text{tot}^\text{surf} r_i m_i/\bar m .
\label{eq:pc}
\end{equation}
By comparing Eq.~\ref{eq:pp} for the surface partial pressure and
Eq.~\ref{eq:pc} for the contribution to the surface pressure, it is clear that
$P_{\text{p},i}^\text{surf} = P_\text{i}^\text{surf}$ only in the special case
where $m_i = \bar m$, i.e. where all atmospheric constituents have the same
molar weight.

As an example, an atmosphere composed of N$_2$ and CO$_2$, with number/volume
mixing ratios $r_\text{N$_2$} = 0.9$ and $r_\text{CO$_2$} = 0.1$ and a total
surface pressure $P_\text{tot}^\text{surf} = \SI{1}{\bar}$, the partial
pressure of each gas at the surface is given by Eq.~\ref{eq:pp}:
\begin{align}
P_{\text{p},\text{N$_2$}}^\text{surf} &= r_\text{N$_2$} P_\text{tot}^\text{surf} = \SI{0.9}{\bar}, \\
P_{\text{p},\text{CO$_2$}}^\text{surf} &= r_\text{CO$_2$} P_\text{tot}^\text{surf} = \SI{0.1}{\bar}.
\end{align}
The mean molecular weight is given by Eq.~\ref{eq:mmw}:
\begin{equation}
\bar m = r_\text{N$_2$} m_\text{N$_2$} + r_\text{CO$_2$} m_\text{CO$_2$} = \SI{29.6}{\gram \per \mole},
\end{equation}
while the contribution of each gas to the total surface pressure is, using Eq.~\ref{eq:pc},
\begin{align}
P_\text{N$_2$}^\text{surf} &= P_\text{tot}^\text{surf} r_\text{N$_2$} m_\text{N$_2$}/\bar m = \SI{0.85}{\bar}, \\
P_\text{CO$_2$}^\text{surf} &= P_\text{tot}^\text{surf} r_\text{CO$_2$} m_\text{CO$_2$}/\bar m = \SI{0.15}{\bar}.
\end{align}
In this example $P_\text{N$_2$}^\text{surf} <
P_{\text{p},\text{N$_2$}}^\text{surf}$ and $P_\text{CO$_2$}^\text{surf} >
P_{\text{p},\text{CO$_2$}}^\text{surf}$. In other words, the contribution of
N$_2$ to the total surface pressure is smaller than its partial pressure at the
surface, while the contribution of CO$_2$ to the total surface pressure is
larger than its surface partial pressure. This may seem counter intuitive, but
is explained by the fact that a part of the partial pressure of each gas at the
surface results from the weight of all gases.

The above discussion shows that giving gas amounts in units of pressure is
ambiguous, and should always be accompanied by a statement specifying if it is
the partial pressures of each gas at the surface or their contributions to the
total surface pressure to avoid confusion. For this reason we prefer to specify
gas amounts in terms of total surface pressure and number/volume mixing ratios,
as this is both unambiguous and also the input required by \rocke{}.

\subsection{Stellar spectra}\label{subsec:StellarSpectra}

When using the default \modele{} radiation scheme, instead of SOCRATES, with
alternative stellar spectra, a software tool is available to bin and format any
high-resolution stellar spectrum for input to the GCM. A Python script is also
provided to plot the outputs, with comparisons to the present day Sun. The
source spectrum should cover the range \SIrange{115}{100000}{\nano \meter}, and
the integral must provide the total stellar flux at any arbitrary known
distance from the stellar surface. The software tool partitions the source
spectrum into a 190-bin spectrum covering this range.  Fluxes outside this
range are integrated over wavelength, divided by the bin size of the end
closest end bin, and added to the flux of this bin. Consequently, the end bins
will have a slightly higher flux to account for the tail fluxes outside the bin
range. The binned output spectrum is utilized for specification of the stellar
constant at the top of the atmosphere of the planet, ultraviolet fluxes to
calculate absorption by ozone, and off-line calculation of another input file
used for photolysis rate estimation. Additionally, a 16-bin array is output
with the fractions of total stellar irradiance per bin for shortwave energy
balance calculations. Detailed instructions and the formatting tool can be
downloaded from the NASA GISS \rocke{} software
webpage.\footnote{http://www.giss.nasa.gov/tools/rocke3d/spectralbin.html}

When using the default \modele{} radiation scheme, users should be aware that
this scheme will have biases with redder stars and may produce significant
errors in fluxes and heating rates for planets with water vapor or other gases
that absorb in the near-infrared. In such cases one would be advised to
consider using the SOCRATES radiation scheme.

\subsection{Topographic Reconstructions}

\begin{figure}
\centering
\includegraphics[scale=0.15]{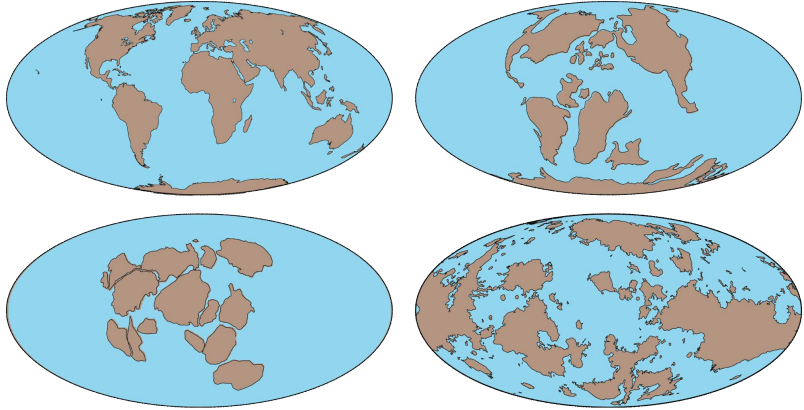}
\caption{Examples of continental configurations utilized in \rocke{}
simulations. Clockwise from upper left: modern Earth, Cretaceous Earth (100
Ma), Neoproterozoic Earth (715 Ma), paleo-Venus with oceans.}
\label{fig:continental_configurations}
\end{figure}

We have made reconstructions of land/ocean distributions and topographic relief
for Earth, Venus \citep{Way2016}, and Mars based on the best available
resolution digital topographies with optional ocean coverage dependent on the
choice of water depth (see Figure \ref{fig:continental_configurations}). 

Reconstructions of paleo-Earth for the past 1 billion years use geologic and
geophysical information, consisting primarily of paleomagnetic data to
determine continental positions, based on reconstructed paleo plate positions.
Following plate tectonic reconstruction sedimentological and paleontological
evidence supplies details about shoreline location, which can differ
substantially from the continental boundaries, depending on the depth of ocean
water. It is worth noting that even minor changes in ocean depth and land/ocean
distribution in the GCM can have significant impacts on the planet's climate
since the opening and closing of ocean gateways, or the orientation of
mountains, can have large impacts on the resulting circulation of the oceans
and atmosphere, and therefore on the transport of heat and moisture.
Topographic relief for paleo-Earth continents is also based on depositional
environment reconstructions using sediment and fossil distribution, but also
considering tectonic settings that arise from plate interactions along
subduction zones, continental rifts, and continent-continent collision zones.
Bathymetry is based on similar evidence, but due to a lack of larger-scale
areas of preserved ocean crust in the geologic record reconstruction of
bathymetry \citep[e.g.][]{Xu2006} for time periods older than 180-200 million
years is not possible.

\section{Post-Processing for Exoplanet Mock Observations} \label{sec:post-processing}

\subsection{Disk-integrated Spectra and Light Curves}

A useful application of the model outputs is the prediction for what the planet
would look like if it were an exoplanet.  A set of external Python codes are
provided for generating the disk-integrated light curves (both reflected and
thermal), as if it were observed from an astronomical distance, given the
externally specified parameters for spin and orbital motion of the planet
(direction of the axes and the periodicity).  The program reads
top-of-atmosphere radiation diagnostics (short-wave and long-wave) from the
model outputs, and integrates the outgoing top-of-atmosphere fluxes in each
radiation band from each pixel over the planetary disk, taking account of the
relative configuration of the planet and the observer.  Given that GCMs are
regularly run with a small number of spectral bands for computational
efficiency, users who want to create a higher resolution spectrum need to run
the model for a short amount of time with a larger number of bands after the
model reaches an equilibrium state.

An isotropic radiation field is assumed in the current scheme.  Thus any
deviation from it, e.g. due to the strongly anisotropic nature of scattering by
clouds, is not included.  The radiation diagnostics used to calculate these
synthetic observations are output as monthly means, consequently the effects of
temporal variation of cloud cover on timescales shorter than the output
frequency are lost.

\begin{figure}
    \centering
    \includegraphics[width=0.5\textwidth]{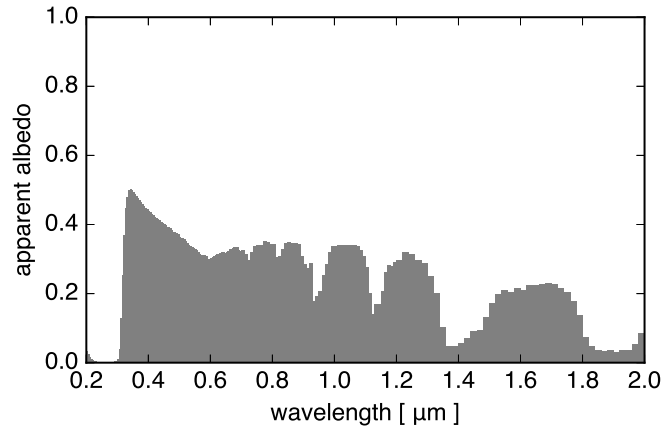}
    \caption{The disk-integrated apparent albedo of the Earth as a function of
wavelength. Based on \rocke{} outputs simulating present day Earth.}
\label{fig:disk_integrated_spectra_SW}
\end{figure}

\begin{figure}
    \centering
    \includegraphics[width=0.5\textwidth]{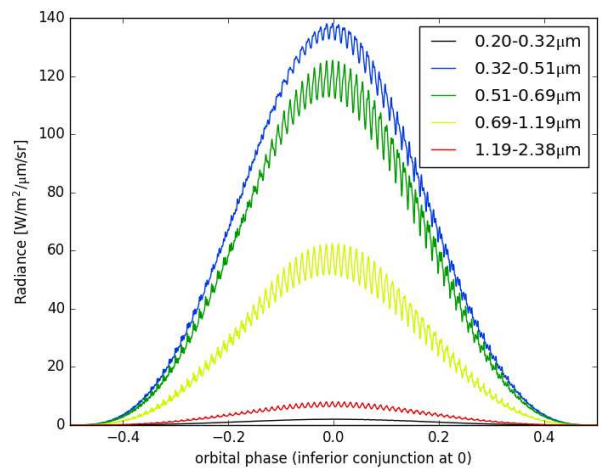}
    \caption{Orbital variation of the disk-integrated scattered light of the
Earth based on \rocke{} outputs simulating  present day Earth. In order to see
the diurnal variations clearly, the spin period is set at 100 hours while the
orbital period is 1 Earth year. Winter solstice for the northern hemisphere is
located at inferior conjunction.} \label{fig:disk_integrated_lightcurves_SW}
\end{figure}

\begin{figure}
    \centering
    \includegraphics[width=0.5\textwidth]{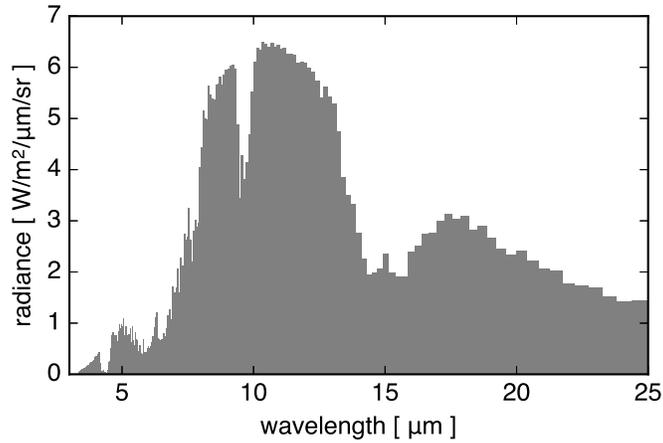}
    \caption{The disk-integrated thermal emission spectrum of the Earth as a
function of wavelength. Based on \rocke{} outputs simulating present day
Earth.} \label{fig:disk_integrated_spectra_LW}
\end{figure}

\begin{figure}
    \centering
    \includegraphics[width=0.5\textwidth]{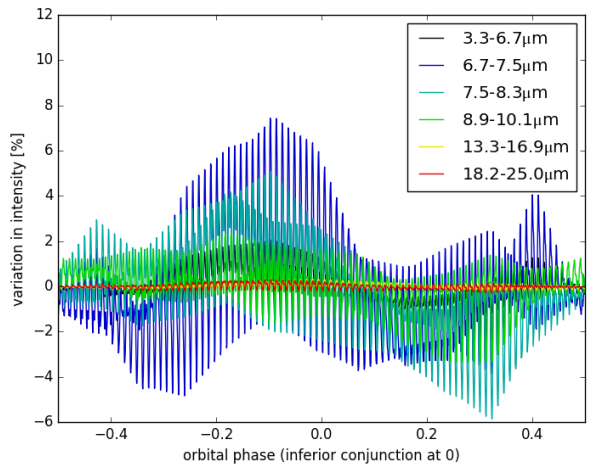}
    \caption{Orbital fractional variation of the disk-integrated thermal
emission of the Earth based on \rocke{} outputs. In order to see the diurnal
variations clearly, the spin period is set at 100 hours, while the orbital
period is 1 Earth year. } \label{fig:disk_integrated_lightcurves_LW}
\end{figure}

Figures~\ref{fig:disk_integrated_spectra_SW}
and~\ref{fig:disk_integrated_spectra_LW} show the annually averaged,
disk-integrated, albedo and thermal emission spectra of the Earth, while
Figs.~\ref{fig:disk_integrated_lightcurves_SW}
and~\ref{fig:disk_integrated_lightcurves_LW} present their variations over one
orbit.  The albedo plotted here is the ``apparent albedo'', defined as the
observed reflected intensity divided by the reflected intensity of a loss-less
Lambert sphere at the same phase.  The light curves assume an orbital
inclination of \ang{90} (i.e. the planet is on an edge-on orbit), \ang{23.4}
obliquity, and that the winter equinox for the northern hemisphere coincides
with inferior conjunction (e.g. the planet is between the star and the
observer).  In order to show the diurnal variations together with the yearly
variations in one panel, the spin period is artificially set to 100 hours while
the orbital period is set at 1 Earth year in Figures
~\ref{fig:disk_integrated_lightcurves_SW}
and~\ref{fig:disk_integrated_lightcurves_LW};  diurnal variations with 24 hour
periodicity would be smeared out when the horizontal axis spans 365 days (see
Figure 3 in \citealt{Fujii2012} for an example of a 24 hour period).  All of
the parameters discussed above can be modified to fit a particular planet.

\subsection{Transmission Spectra}

A second set of external Python codes compute the transmission spectrum of a
planet based on output from \rocke{}.  The code reads the atmospheric columns
located near the terminator, interpolates them onto the terminator, and
computes the transmission of the starlight based on these profiles.  Examples
of required input parameters are; location of the GCM outputs, the composition
of the background atmosphere, the cross-section tables to use, wavelength
resolution, the radius of the star and the planet, and the viewing geometry of
the star, planet, and the observer.

\begin{figure}
    \centering
    \includegraphics[width=0.5\textwidth]{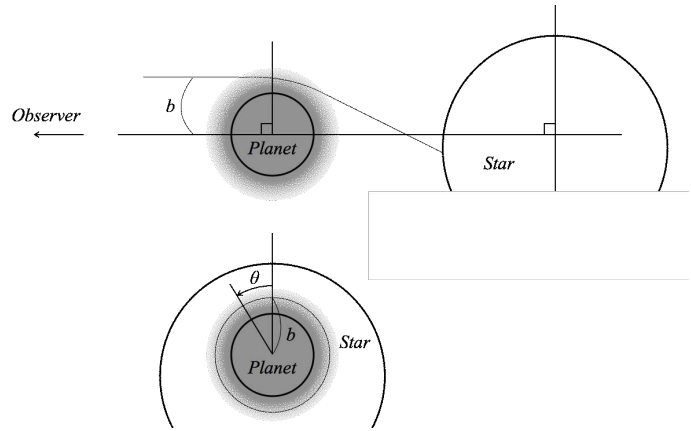}
    \caption{Geometry of transmission spectroscopy.}
    \label{fig:transmission_geometry}
\end{figure}

We consider polar coordinates $(b, \theta )$ on the spherical plane centered at
the planetary center, as illustrated in Fig.~\ref{fig:transmission_geometry}.
Denote the spectral extinction optical depth along the ray that exits the
planetary atmosphere at $(b, \theta )$ by $\tau (\lambda; b, \theta )$, the
fraction of the intensity absorbed or scattered along the optical path that
exits the atmosphere at the altitude between $b$ and $b+db$, and at the angle
between $\theta $ and $\theta + d\theta $, $f(b, \theta ) \, d\theta \,db$, is
then
\begin{equation}
f(\lambda; b, \theta ) \, d\theta \, db = ( 1 - e^{-\tau (\lambda; b, \theta )} ) \, b \, d\theta \, db .
\label{eq:transmission}
\end{equation}
Using $f(b, \theta )$, the transit depth, $\Delta F$, is given by
\begin{equation}
\pi R_{\star } ^2 \, \Delta F (\lambda) = \pi b_{\rm min}^2 + \int_0^{2\pi} d\theta \int_{b_{\rm min}}^{b_{\rm max}} db \, b\, f(\lambda; b, \theta ),
\end{equation}
where $b_{\rm min}$ is the smallest impact parameter at which the ray is
completely attenuated, and $b_{\rm max}$ is the impact parameter at which the
planetary atmosphere may be regarded as transparent.  We assume $b_{\rm max} =
R_{\rm p} + \SI{100}{\kilo \meter}$, where $R_{\rm p}$ is the planetary radius,
which is large enough to cover the whole vertical domain included in \rocke{}.
Transmission spectra are also regularly represented by an ``effective height'',
$h_{\rm eff}$, or ``effective radius'', $R_{\rm p}^{\rm eff} = R_{\rm p} +
h_{\rm eff}$, which is given by
\begin{equation}
\Delta F(\lambda) = \frac{R_{\rm p}^{\rm eff}(\lambda)^2}{R_{\star }^2} = \frac{( R_{\rm p} + h_{\rm eff}(\lambda)  )^2}{R_{\star }^2}.
\end{equation}

The trajectory of the transmitted ray seen by the observer is traced from the
direction of the observer toward the stellar disk, accounting for the
refraction due to the planetary atmosphere as described in \citet{Misra2014}
and \citet{vanderWerf2008}.  The index of refraction of the atmosphere is
computed based on the number density of the atmosphere's constituents.  If the
ray that transverses at a certain altitude of the atmosphere does not intersect
the stellar disk it is observed as opaque \citep{Betremieux2014,Misra2014}.

The opacities along the ray trajectories, $\tau (\lambda; b, \theta )$ in
equation (\ref{eq:transmission}), are computed and take into account both
gaseous Rayleigh scattering and absorption.  The cross-sections of atmospheric
molecules are calculated based on HITRAN~2012 \citep{Rothman2013}, with both
Doppler broadening and pressure broadening by air accounted for in the Voigt
profile of each line.  The cross section data are tabulated at temperatures
between \SIlist{100;400}{\kelvin} in steps of \SI{50}{\kelvin} and pressures
(equally spaced logarithmically in intervals of 1-order of magnitude in
millibar units), that will be interpolated to obtain the absorption coefficient
at each location along the trajectory.  The cross section tables for O$_2$,
O$_3$, H$_2$O, CO$_2$, CH$_4$, N$_2$O are provided, while those for other
molecules can be created from the corresponding HITRAN data with the provided
codes.  The opacity due to cloud particles may be included in a simplified
manner. However, GCM outputs of the cloud properties are typically averaged
over one month which is significantly longer than the time scale of cloud
formation, but it is possible to obtain averages over intervals as small as 30
minutes using the \modele{} Sub-daily (SUBDD) facility.\footnote{See Part II
section 3 of the \modele{} on-line user guide
\url{https://simplex.giss.nasa.gov/gcm/doc/UserGuide/diagnostics.html}}

\begin{figure}
    \centering
    \includegraphics[width=0.6\textwidth]{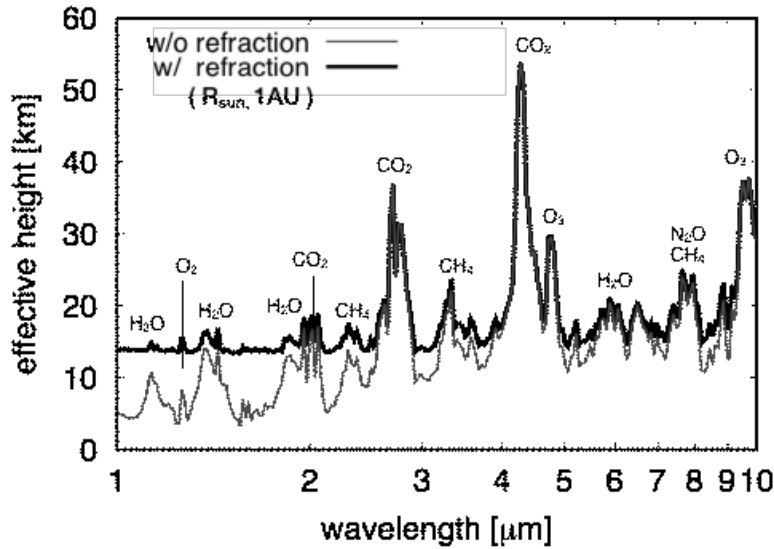}
    \caption{Cloud-free transmission spectra based on \rocke{} experiments
simulating the present day Earth. Results are shown both with the effect of
refraction (black thick line) and without (grey, thin line). A stellar radius
of $R_\text{sun}$ and a planetary orbital distance of \SI{1}{\astronomicalunit}
were used.} \label{fig:transmission}
\end{figure}

We note that the transmittance based on the time-averaged optical properties
(which we calculate) may be somewhat different from the time-averaged
transmittance based on the instantaneous optical properties (which is
observed).  Figure~\ref{fig:transmission} displays examples of the cloud-free
transmission spectra based on an Earth GCM simulation with and without the
effect of refraction in the planetary atmosphere.  In demonstrating the effect
of refraction, we assumed that the radius of the host star was $1 R_{\rm sun}$
and the planet's orbital distance was \SI{1}{\astronomicalunit}, and the
planetary center coincided with the center of the stellar disk. 
These assumed geometrical parameters can
easily be modified to fit any particular planet.  The results shown in
Fig.~\ref{fig:transmission} are consistent with previous literature
results~\citep[e.g.][]{Kaltenegger2009, Betremieux2014, Misra2014}.

\newpage

\section*{Acknowledgments}

We thank James Manners for providing access to SOCRATES and advice on coupling
it to \rocke{}. This research was supported by the NASA Astrobiology Program
through our participation in the Nexus for Exoplanet System Science, and by the
NASA Planetary Atmospheres Program, Exobiology Program, and Habitable Worlds
Program.  We also acknowledge internal Goddard Space Flight Center Science Task
Group funding that triggered the initial development of \rocke{} along with
help from Shawn Domagal-Goldman. Resources supporting this work were provided
by the NASA High-End Computing (HEC) Program through the NASA Center for
Climate Simulation (NCCS) at Goddard Space Flight Center. The Viking 2 lander
VL1/VL2-M-MET-4-BINNED-P-T-V-V1.0 data set was obtained from the NASA Planetary
Data System.

\bibliographystyle{apj}
\bibliography{bibliography}

\end{document}